\documentclass[12pt]{article}
\pdfoutput=1
\usepackage[dvipsnames]{xcolor}
\usepackage{cite}
\usepackage{hyperref}
\usepackage{subcaption}
\usepackage{bookmark}
\usepackage{setspace}
\usepackage{braket}
\usepackage{amsmath, amssymb, amsthm, float, graphicx}
\numberwithin{equation}{section}

\hypersetup{colorlinks=false, linkcolor=blue, citecolor=red}

\def\beq{\begin{eqnarray}}\def\eeq{\end{eqnarray}}
\def\be{\begin{equation}}\def\ee{\end{equation}}
\def\bs{\begin{split}}\def\es{\end{split}}

\def\m{\mu}

\def\a{\alpha}

\def\b{\beta}
\def\d{\delta}

\def\l{\lambda}

\def\wmW{{\widehat{\mathcal{W}}}}
\def\mW{{\mathcal{W}}}
\def\zt{\tilde{z}}

\def\immpGo{\text{Disc}_s\left[\mathcal{G}_0\left(s_{1}^{\prime} ; s_{2}^{(+)}\left(s_{1}^{\prime}, a\right)\right)\right]}
\def\immpG1{\text{Disc}_s\left[\mathcal{G}_1\left(s_{1}^{\prime} ; s_{2}^{(+)}\left(s_{1}^{\prime}, a\right)\right)\right]}
\def\immpG2{\text{Disc}_s\left[\mathcal{G}_2\left(s_{1}^{\prime} ; s_{2}^{(+)}\left(s_{1}^{\prime}, a\right)\right)\right]}

\usepackage{amstext} 
\usepackage{array}   
\newcolumntype{L}{>{$}l<{$}} 

\usepackage[utf8]{inputenc}

\usepackage{lipsum}

\begin{document}

\title{\bf Positivity and Geometric Function Theory Constraints on Pion Scattering}
\date{}
\author{Ahmadullah Zahed\\
\it Centre for High Energy Physics,
\it Indian Institute of Science,\\ \it C.V. Raman Avenue, Bangalore 560012, India. \\
\href{mailto:ahmadullah@iisc.ac.in}{\texttt{ahmadullah@iisc.ac.in 
}}}
\maketitle
\vskip 2cm
\abstract{This paper presents the fascinating correspondence between the geometric function theory and the scattering amplitudes with $O(N)$ global symmetry. A crucial ingredient to show such correspondence is a fully crossing symmetric dispersion relation in the $z$-variable, rather than the fixed channel dispersion relation. We have written down fully crossing symmetric dispersion relation for $O(N)$ model in $z$-variable for three independent combinations of isospin amplitudes. We have presented three independent sum rules or locality constraints for the $O(N)$ model arising from the fully crossing symmetric dispersion relations. We have derived three sets of positivity conditions. We have obtained two-sided bounds on Taylor coefficients of physical Pion amplitudes around the crossing symmetric point (for example, $\pi^+\pi^-\to \pi^0\pi^0$) applying the positivity conditions and the Bieberbach-Rogosinski inequalities from geometric function theory.}

\tableofcontents

\onehalfspacing

\section{Introduction}
In geometric function theory, Bieberbach-Rogosinski inequalities tell us that the Taylor coefficients of typically real functions (also known as Herglotz function) are two-sided bounded \cite{rogo, komatu}. A typically real function $f(z)$,  is a function that satisfies 
$\text{Im}[f(z)] \text{Im}[z] > 0, ~~ \text{Im}[z]\neq 0$, \textit{i.e.}  the  imaginary part of the function is positive in the upper half plane, and negative in the lower half plane. Let $f(z)$ be a typically real function with (normalized) Taylor expansion (inside the disk $|z|<1$)
\be
f(z)=z+\sum_{p=2}^{\infty}b_p z^p\,.
\ee
Then the Bieberbach-Rogosinski inequalities say
\be\label{eq:Rogbounds}
-\kappa_p \leq b_p \leq p \,,
\ee
where
\be\nonumber
\kappa_{p}=-\text{Min}\left(\frac{\sin p \theta}{\sin \theta}\right)\,,~ 0\leq \theta \leq \pi.
\ee
which is given by 
\be
\kappa_{p}=-\frac{\sin p \varphi_{p}}{\sin \varphi_{p}}\,,\,\,\csc \left(\varphi _p\right) \left(p \cos \left(p \varphi _p\right)-\sin \left(p \varphi _p\right) \cot \left(\varphi _p\right)\right)=0 
\ee
For  $p$ odd, $\pi/p<\varphi_{p}<3\pi/2p$, while for $p$ even, $\varphi_{p}=\pi$ is only solution, giving $\kappa_p = p$.

For $p$ odd $\kappa_p $ given by solutions to
the above equation \cite{PRAS, rogo, komatu}. For examples
$\kappa_3 = 1, \kappa_5 \approx 1.25, \kappa_7 \approx 1.6 ,\dots$.\\

In \cite{PHASAZ} the correspondence between the famous Bieberbach conjecture (de Branges’ theorem) and the non-perturbative crossing symmetric scattering amplitudes was pointed out, which established a close relationship between the Bieberbach-bounds and the bounds on the Wilson coefficients. In \cite{PRAS} it was pointed out that crossing symmetric scattering amplitudes are typically real functions. The pivotal ingredient in demonstrating such a correlation between the geometric function theory and the scattering amplitudes is crossing symmetric dispersion relation in a new $z$-variable for a fixed parameter $a$. The $z,a$ variables arise by parametrizing the Mandelstam invariants in the following way. The 2-2 scattering amplitudes with $O(N)$ global symmetry are functions of $s, t, u$, the Mandelstam invariants, satisfying $s + t + u = 4m^2$, where $m$ is the mass of the external scalars. For convenience, we will work with the notation \cite{AK} \cite{ASAZ} 

\be 
s_{1}=s-\frac{\mu}{3},~ s_{2}=t-\frac{\mu}{3},~ s_{3}=u-\frac{\mu}{3}=-s_1-s_2,~\mu=4m^2;
\ee
Fully crossing symmetric dispersion relation is written down by parametrizing $s_i$ as a function of $z,a$. 
The parametrization is given by
\be
s_{i}=a-\frac{a\left(z-z_{i}\right)^{3}}{z^{3}-1}, \quad i=1,2,3\,,
\ee
where $a$ is a real parameter, $-\frac{\m}{3}\leq a<\frac{2\m}{3}$ and $z_i$ is the cube roots of unity, after we parametrize $s_i$ as a function of $z,a$, the amplitude is a function of $z^3=\zt,a$,\textit{i.e.} $\overline{\mathcal{M}}(\zt,a)$. The parameter $a$ is given by $a=\frac{y}{x}$, where $x=-\left(s_1 s_2 + s_2 s_3+s_3 s_1\right)$, $y=-s_1 s_2 s_3$. For a crossing symmetric dispersion relation in $z$-variable, while keeping $a$-fixed \cite{AK,ASAZ}, the kernel is a univalent function\footnote{A function is univalent on a domain $\mathcal{D}$ if it is holomorphic, and one-to-one, i.e. for all $z_1, z_2$ in $\mathcal{D}$, $f(z_1)= f(z_2)$ if
$z_1= z_2$.}, and the absorptive part is positive for a certain range of $a$. The kernel's specific form and the absorptive part's positivity enable one to re-express the dispersion relation as a Robertson representation\footnote{\textit{In $|z| < 1$ a regular function $F(z)$ is typically
real if and only if it has the Robertson representation:
\be
F(z)=\int_{-1}^{1} d f(\eta) \frac{z}{1-2 \eta z+z^{2}}\,,
\ee
where the measure $f(\eta)$ is a non-decreasing function.}}. Once a function has a Robertson representation, we can say it is a typically real function. Therefore the amplitudes $\overline{\mathcal{M}}(\zt,a)$ are typically real functions. We ask here if such dispersion relation can be written for theories with $O(N)$ global symmetry? Do these correspondences exist for theories with global $O(N)$ symmetry? The answer turns out to be yes! We write down three sets of fully crossing symmetric dispersion relations for three specific combinations of isospin amplitudes. We call these combinations of isospin amplitudes to be $\mathcal{F}_k(\zt,a),~~k=0,1,2$. We find that these combinations are in the Robertson representation. Hence we can write \textit{three sets of Bieberbach-Rogosinski inequalities}. If we write
\be
\mathcal{F}_k(\zt,a)=\sum_{p=1}^{\infty}\a_p^{(k)}(a)a^{2p}\zt^{p},~~k=0,1,2\,,
\ee
the Bieberbach-Rogosinski inequalities take the form:
\be
-\kappa_p \leq \frac{\a_p^{(k)}(a)a^{2p}}{\a_1^{(k)}(a)a^{2}}\leq p\,,
\ee
for a range of $a$, which is derived in the main text (see eq \eqref{eq:Rogbounds}).

These three sets of dispersion relation and unitarity conditions give \textit{three sets of positivity constraints} for Taylor coefficients of the amplitudes around the crossing symmetric point. These three sets of new positivity constraints are very non-trivial. Demanding the locality, we get \textit{three sets of independent sum rules or locality constraints}. These novel sets of independent sum rules constrain the theories strongly.

There is a lot of recent work on constraining the quantum field theory using dispersion relation, utilizing the analyticity and unitarity assumptions \cite{mandelstam,nuss,Martin,Meltzer,nima}. Dispersion relations are the non-perturbative depiction of scattering amplitudes. Usually, people write dispersion relations in 2-2 scattering by keeping Mandelstam invariant $t$-fixed and write a dispersion integral in $s$-variable, which leads to the $s\leftrightarrow u$ symmetric representation of the amplitude. Imposing full crossing symmetry as an additional condition, one gets null constraints. Analogous strategies also developed in the context of conformal field theories, Mellin amplitudes.  See for example \cite{joaopaper,joaopaper2,cmrs,RGASAZ}.

The constraints in EFT Wilson coefficients were worked out in \cite{RMTZ,rattazzi,TWZ,Wang:2020jxr,chout,chout2,
chout3,nimayutin,green,nimayutin2,yutin2,sasha,Davis:2021oce}, using positivity of the partial wave and null conditions\footnote{Locality constraints are the same as the null constraints. See \cite{ASAZ} for more discussion}. In our case, we don’t use the null constraints; instead, we use positivity (of the partial wave expansion) and bounds in the Taylor series coefficients of the amplitudes in $z$-variable that appear from the geometric function theory. 

\textit{Bieberbach-Rogosinski inequalities and the positivity conditions give two-sided bounds on the Taylor coefficients of the amplitudes of any physical process around the crossing symmetric point}\footnote{The Bieberbach-Rogosinski inequalities and the positivity conditions are insufficient to put two-sided bounds for solely two-channel symmetric amplitudes without global symmetry. There non-linear inequalities and null constraints are needed. See \cite{PRAS} for further details. We thank Prashanth Raman for pointing out this to us.}; these processes need not be fully crossing symmetric, only two channel symmetry is sufficient. For example, consider
\be
\mathcal{M}^{(\pi^+\pi^-\to \pi^0\pi^0)}(s_1,s_2)=\sum_{p=0,q=0}^{\infty}\mathcal{C}_{p,q}(-s_2 s_3)^p (-s_1 )^q\,.
\ee
\\
This amplitude is not fully crossing symmetric. The three sets of Bieberbach-Rogosinski inequalities and three sets of positivity conditions provide two-sided (order $\mathcal{O}(1)$, in \cite{PRAS} the $\mathcal{O}(1)$ is shown to be $|\mathcal{O}(1)|<5.625$) bounds on the $\mathcal{C}_{p,q}$, presented in table \eqref{tab:Cpqbounds}.

We have organized the paper in the following way. The definitions of the fully crossing symmetric combinations, their dispersion relations, inversion formula and sum rules are presented in Section 2. Section 3 contains positivity conditions on the Taylor coefficients of three crossing symmetric combinations. Section 4 describes how geometric function theory for the $O(N)$ model can be realized. Section 5 contains the application of the geometric function theory to physical pion amplitudes and bounds on Taylor coefficients of physical amplitudes around crossing symmetric point. We conclude with a summary and future directions in section 6. We have added Appendices for multiple demonstrations and verifications.

\section{Crossing symmetric dispersion relation for $O(N)$ model}
The 2-2 scattering amplitude with $O(N)$ global symmetry can be written as
\be
\begin{aligned}
\mathcal{M}_{a b}^{c d}(s_1,s_2)=& A(s_1 \mid s_2, s_3) \delta_{a b} \delta^{c d}+A(s_2 \mid s_3, s_1) \delta_{a}^c \delta_{b }^d+A(s_3 \mid s_1, s_2) \delta_{a}^d \delta_{b }^c=\sum_{I=0}^{2}\mathcal{M}^{(I)}(s_1,s_2)\mathbb{P}_I\,,
\end{aligned}
\ee
where $A(s_i \mid s_j, s_k)=A(s_i \mid s_k, s_j)$. The isospin $I$ $s$-channel amplitudes $\mathcal{M}^{(I)}(s_1,s_2)$ are given by
\be\label{eq:isoM}
\begin{split}
&\mathcal{M}^{(0)}(s_1,s_2)=N A(s_1 \mid s_2, s_3)+A(s_2 \mid s_3, s_1)+A(s_3 \mid s_1, s_2) \,, \\
&\mathcal{M}^{(1)}(s_1,s_2)=A(s_2 \mid s_3, s_1)-A(s_3 \mid s_1, s_2)\,,\\
& \mathcal{M}^{(2)}(s_1,s_2)=A(s_2 \mid s_3, s_1)+A(s_3 \mid s_1, s_2) \,,
\end{split}
\ee
with
\be
\mathbb{P}_{0}=\frac{1}{N} \delta_{a b} \delta^{c d}, \quad \mathbb{P}_{1}=\frac{1}{2}\left(\delta_{a}^{c} \delta_{b}^{d}-\delta_{a}^{d} \delta_{b}^{c}\right), \quad \mathbb{P}_{2}=\frac{1}{2}\left(\delta_{a}^{c} \delta_{b}^{d}+\delta_{a}^{d} \delta_{b}^{c}-\frac{2}{N} \delta_{a b} \delta^{\mathrm{cd}}\right)\,.
\ee
We want to write down crossing symmetric dispersion relation in $z$-variable for fixed $a$ following \cite{AK, ASAZ} for $O(N)$ model. We are interested in such a crossing symmetric dispersion relation to connecting with the geometric function theory. In general full 3-channel crossing symmetry is missing in isospin amplitudes $\mathcal{M}^{(I)}(s_1,s_2)$. Following \cite{roy, Roskies:1970uj}, we will consider independent fully crossing symmetric combinations 
\be\label{def:G0}
\mathcal{G}_0(s_1,s_2)=\frac{\mathcal{M}^{(0)}\left(s_1,s_2\right)}{N}+\frac{(N-1) \mathcal{M}^{(2)}\left(s_1,s_2\right)}{N}\,,
\ee
\be\label{def:G1}
\mathcal{G}_1(s_1,s_2)=\frac{\mathcal{M}^{(1)}\left(s_3,s_1\right)}{s_1-s_2}+\frac{\mathcal{M}^{(1)}\left(s_1,s_2\right)}{s_2-s_3}+\frac{\mathcal{M}^{(1)}\left(s_2,s_3\right)}{s_3-s_1}\,,
\ee
\be\label{def:G2}
\mathcal{G}_2(s_1,s_2)=\frac{\frac{\mathcal{M}^{(1)}\left(s_1,s_2\right)}{s_2-s_3}+\frac{\mathcal{M}^{(1)}\left(s_2,s_1\right)}{s_3-s_1}}{s_1-s_2}+\frac{\frac{\mathcal{M}^{(1)}\left(s_2,s_3\right)}{s_3-s_1}+\frac{\mathcal{M}^{(1)}\left(s_3,s_2\right)}{s_1-s_2}}{s_2-s_3}+\frac{\frac{\mathcal{M}^{(1)}\left(s_1,s_3\right)}{s_2-s_3}+\frac{\mathcal{M}^{(1)}\left(s_3,s_1\right)}{s_1-s_2}}{s_3-s_1}\,.
\ee
The $\mathcal{G}_0(s_1,s_2)$ is the $\pi^0\pi^0\to\pi^0\pi^0$ amplitude. Dispersion relation in $z$-variable for $\mathcal{G}_0(s_1,s_2)$ is discussed in \cite{AK, ASAZ}, and relations to geometric function theory is given in detail in \cite{PHASAZ, PRAS}. We will write $z$-variable dispersion relations for all three of the $\mathcal{G}_k(s_1,s_2)$ and  demonstrate that $\mathcal{G}_k(s_1,s_2)$ (linear combinations of them) are related to geometric function theory. We expect that $\mathcal{G}_k(s_1,s_2)$ to have the same analyticity properties as the isospin amplitude $\mathcal{M}^{(I)}(s_1,s_2)$. Since antisymmetry of $\mathcal{M}^{(1)}(s_1,s_2)$ with respect to $s_2$ and $s_3$, prevents the denominators appearing in $\mathcal{G}_1(s_1,s_2)$ and $\mathcal{G}_2(s_1,s_2)$ to introduce any additional new singularities at $s_1 = s_2,~s_2 =s_3, ~s_3 = s_1$. From the large $s_1$, fixed $s_2$ behaviour of the isospin amplitudes $\mathcal{M}^{(I)}(s_1,s_2)$, we can write once\footnote{For $\mathcal{G}_2(s_1,s_2)$ we can write a dispersion relation without subtraction, which we will not use here.} subtracted dispersion relation in $z$-variable for fixed $a$ for $\mathcal{G}_k(s_1,s_2)$. For large $s_1$ fixed $s_2$ we have
\be
\mathcal{G}_k(s_1,s_2)=o(s_1^2),~\text{for fixed } s_2\,.
\ee
The discontinuity of $\mathcal{G}_k(s_1,s_2)$ starts at $s_1=\frac{2\m}{3}$. Therefore, once we know the discontinuity of $\mathcal{G}_k(s_1,s_2)$, we can write fully crossing symmetric dispersion relations, following the same logic in \cite{AK,ASAZ}. This leads to fully crossing symmetric dispersion relation\footnote{Crossing symmetric dispersion relation in \cite{roy} are in $x,y$ variables with a complicated kernel, not in $z$-variable. It is not clear if, from the dispersion relation in \cite{roy}, one can connect with geometric function theory.} is $s_1,s_2$ variables  
\be
\label{eq:disper00}
\mathcal{G}_{k}(s_1, s_2)=\alpha_{0}^{(k)}+\frac{1}{\pi} \int_{\frac{2\m}{3} }^{\infty} \frac{d s_1^{\prime}}{s_1^{\prime}} \text{Disc}\mathcal{G}_{k}\left(s_1^{\prime} ; s_2^{(+)}\left(s_1^{\prime} ,a\right)\right) H\left(s_1^{\prime} ;s_1, s_2, s_3\right)\,;\,\, k=0,1,2\,,
\ee
where $\a_0^{(k)}$ is the subtraction constant. The $\text{Disc}\mathcal{G}_{k}\left(s_1; s_2\right)$ is the $s$-channel discontinuity of $\mathcal{G}_{k}(s_1, s_2)$ and
\be
\begin{split}
&H\left(s_1^{\prime} ; s_1, s_2,s_3\right)=\left[s_1\left(s_1^{\prime}-s_1\right)^{-1}+s_2\left(s_1^{\prime}-s_2\right)^{-1}+s_3\left(s_1^{\prime}-s_3\right)^{-1}\right]\,,\\
&s_{2}^{(+)}\left(s_1^{\prime}, a\right)=-\frac{s_1^{\prime}}{2}\left[1 - \left(\frac{s_1^{\prime}+3 a}{s_1^{\prime}-a}\right)^{1 / 2}\right],~a=\frac{s_1 s_2 s_3}{s_1s_2+s_2s_3+s_3s_1}\,.
\end{split}
\ee
Partial wave expansion for $\mathcal{G}_k(s_1,s_2)$ and it's convergence has been discussed in \cite{roy} (see \cite{anant} for some applications). Since the $\mathcal{G}_k(s_1,s_2)$ have the same analyticity properties as the isospin amplitude $\mathcal{M}^{(I)}(s_1,s_2)$ and $\mathcal{M}^{(I)}(s_1,s_2)$ has cuts starting at $s_1\geq 2\mu/3$, hence the domain of $a$ is same as discussed in \cite{AK}. The range of $a$, we will consider here is  $-\frac{\m}{3}\leq a<\frac{2\m}{3}$. This range of $a$ can be enlarged to $-6.71\mu\leq a<\frac{2\m}{3}$ as discussed in \cite[see appendix ]{AK}, \cite[see discussion below eq. 3.10]{roy}. We will use that in our calculations in the upcoming sections.

Below we have presented the the formula for $s$-channel discontinuity, $ \text{Disc}\mathcal{G}_{k}\left(s_1; s_2\right)$:
\be\label{eq:DiscG_k}
\begin{split}
\text{Disc}\mathcal{G}_{0}\left(s_1; s_2\right)&=\frac{\mathcal{A}^{\text{(0)}}\left(s_1,s_2\right)+(N-1) \mathcal{A}^{\text{(2)}}\left(s_1,s_2\right)}{N}\,,\\
\text{Disc}\mathcal{G}_{1}\left(s_1; s_2\right)&=\frac{3 s_1 \left(s_1+2 s_2\right)\left(2 \mathcal{A}^{\text{(0)}}\left(s_1,s_2\right)- (N+2) \right) \mathcal{A}^{\text{(2)}}\left(s_1,s_2\right)+3 N \left(s_1^2-2 s_2 s_1-2 s_2^2\right) \mathcal{A}^{\text{(1)}}\left(s_1,s_2\right)}{2 N \left(s_1-s_2\right) \left(2 s_1+s_2\right) \left(s_1+2 s_2\right)}\,,\\
\text{Disc}\mathcal{G}_{2}\left(s_1; s_2\right)&=\frac{3 \left(( s_1+2s_2)\left(-2 \mathcal{A}^{\text{(0)}}\left(s_1,s_2\right)+(N+2) \right) \mathcal{A}^{\text{(2)}}\left(s_1,s_2\right)+3 N s_1 \mathcal{A}^{\text{(1)}}\left(s_1,s_2\right)\right)}{2 N \left(s_1-s_2\right) \left(2 s_1+s_2\right) \left(s_1+2 s_2\right)}\,,
\end{split}
\ee
where, $\mathcal{A}^{(I)}(s_1,s_2)$ is the $s$-channel discontinuity of the isospin amplitude $\mathcal{M}^{(I)}(s_1,s_2)$. The $s$-channel discontinuity of the isospin amplitude, $\mathcal{A}^{(I)}(s_1,s_2)$ has a partial wave expansion
\be\label{eq:IsoA_wave}
\begin{split}
\mathcal{A}^{(I)}\left(s_1,s_2^{(+)}(s_1,a)\right)&=\Phi(s_1;\a)\sum_{\ell=0}^{\infty}\left(2\ell+2\a\right)~a_\ell^{(I)}(s_1)~C^{(\a)}_{\ell}\left(\sqrt{\xi(s_1,a)}\right)\,,\\
\xi(s_1,a) &=\cos ^{2} \theta_{s}=\left(1+\frac{2 s_2^{+}(s_1,a)+\frac{2\mu}{3}}{s_1-\frac{2\mu}{3}}\right)^{2}=\xi_0+4\xi_0\left(\frac{a}{s_1-a} \right)\,.
\end{split}
\ee
We normalize $~C^{(\a)}_{\ell}$ such that $\Phi(s_1,\a)=\Psi(\alpha) \frac{\sqrt{s_{1}+\frac{\mu}{3}}}{\left(s_{1}-\frac{2 \mu}{3}\right)^{\alpha}}$ with $\Psi(\alpha) >0$ \textit{i.e} $\Psi(\alpha)$ is a real positive number. More importantly, due to unitarity, the partial wave coefficients satisfy
\be\label{eq:positiveaell}
0\leq a_\ell^{(I)}(s_1)\leq 1 \,.
\ee
For our calculations, we will be only utilizing the positivity of the partial wave coefficients. 
\\
The isospin amplitudes can be written in terms of $\mathcal{G}_k(s_1,s_2)$ in the following way
\be\label{eq:isotoGk}
\begin{array}{l}
 \mathcal{M}^{\text{(0)}}\left(s_1,s_2\right)=\frac{1}{9} \left(3 (N+2) \mathcal{G}_0\left(s_1,s_2\right)+(N-1) \left(3 s_1 \mathcal{G}_1\left(s_1,s_2\right)+\left(-s_1^2+2 s_2 s_1+2 s_2^2\right) \mathcal{G}_2\left(s_1,s_2\right)\right)\right)\,,\\
 \mathcal{M}^{\text{(1)}}\left(s_1,s_2\right)=\frac{1}{3} \left(s_1+2 s_2\right) \left(\mathcal{G}_1\left(s_1,s_2\right)+s_1 \mathcal{G}_2\left(s_1,s_2\right)\right)\,, \\
 \mathcal{M}^{\text{(2)}}\left(s_1,s_2\right)=\frac{1}{9} \left(6 \mathcal{G}_0\left(s_1,s_2\right)-3 s_1 \mathcal{G}_1\left(s_1,s_2\right)+\left(s_1^2-2 s_2 s_1-2 s_2^2\right) \mathcal{G}_2\left(s_1,s_2\right)\right)\,. \\
\end{array}
\ee

These three equations are crucial. \textit{Once the Taylor coefficients of the expansion of $\mathcal{G}_k(s_1,s_2)$ are bounded (see equation \eqref{eq:GktoWk} and table \eqref{tab:boundsWk}), utilizing the above three formulae \eqref{eq:isotoGk}, we can bound the Taylor expansion (around crossing symmetric point) coefficients of amplitude of any physical process .}

\subsection*{Inversion formulas and sum rules}
The $\mathcal{G}_k(s_1,s_2)$ have the same analyticity properties as the isospin amplitude $\mathcal{M}^{(I)}(s_1,s_2)$, and they don't have any additional singularities. More importantly, all three of them are fully crossing symmetric. Therefore, we can write 
\be
\label{eq:GktoWk}
\mathcal{G}_{k}(s_1,s_2)=\sum_{p, q=0}^{\infty} {\mathcal W}_{p q}^{(k)} x^{p} y^{q}\,;\,\, k=0,1,2\,.
\ee
with crossing symmetric variables $x=-\left(s_1 s_2 + s_2 s_3+s_3 s_1\right)$, $y=-s_1 s_2 s_3$. In the $z$-variable the kernel takes the form $H\left(s_1^{\prime} ; s_1, s_2,s_3\right)=\frac{27 a^2 z^3 \left(3 a-2 s_1'\right)}{-27 a^3 z^3+27 a^2 z^3 s_1'+\left(z^3-1\right)^2 \left(s_1'\right){}^3}$, which can be seen by writing $s_i$'s in terms of $(z,a)$. Now identifying crossing symmetric variable $x$ in terms of $z$-variable via the relation $\frac{z^{3}}{\left(z^{3}-1\right)^{2}}=\frac{-x}{27a^2}$, we can series expand in powers of $x$.
We obtain 
\be\label{eq:G_k_ax}
\mathcal{G}_k(s_1,s_2)=\alpha_0^{(k)}+\sum_{n=1}^{\infty} \frac{1}{\pi} \int_{2\m / 3}^{\infty} d s_1' \frac{\text{Disc}\mathcal{G}_{k}\left(s_1^{\prime} ; s_2^{(+)}\left(s_1^{\prime} ,a\right)\right)}{s_1^{' 2 n+1}}\left(1-\frac{a}{s_1'}\right)^{n-1}\left(2-\frac{3 a}{s_1'}\right) x^{n}\,.
\ee
Coefficient of $a^m$ (since $a=y/x$) will be $\mW_{n-m,m}$. In general, one can write
\be\label{eq:invWk}
\mW^{(k)}_{n-m,m}=\int_{2\m / 3}^{\infty} \frac{d s_1}{s_1} ~\Phi(s_1;\a)\sum_{\ell=0}^{\infty}\left(2\ell+2\a\right)~\mathcal{B}_{n,m,\ell}^{(k)}(s_1)\,,~n\geq 1\,.
\ee
\\
The $\text{Disc}\mathcal{G}_{k}$ are given in terms of the partial wave coefficients of the $s$-channel discontinuity of the isospin amplitudes,
\be\label{eq:DiscG_k_aI}
\begin{split}
\text{Disc}\mathcal{G}_{0}(s_1,s_2^{(+)}(s_1,a))&=\Phi(s_1;\a)\sum_{\ell=0}^{\infty}\left(2\ell+2\a\right)~\left(\frac{a_\ell^{(0)}(s_1)}{N}+\frac{(N-1)a_\ell^{(2)}(s_1)}{N}\right)~C^{(\a)}_{\ell}\left(\sqrt{\xi(s_1,a)}\right)\,,\\
\text{Disc}\mathcal{G}_{1}(s_1,s_2^{(+)}(s_1,a))&=\Phi(s_1;\a)\sum_{\ell=0}^{\infty}\left(2\ell+2\a\right)~3 \Bigg(\frac{\left(s_1-a\right) a_{\ell }^{\text{(0)}}\left(s_1\right)}{N s_1 \left(2 s_1-3 a\right)}+\frac{\sqrt{s_1-a} \left(s_1-3 a\right) a_{\ell }^{\text{(1)}}\left(s_1\right)}{2 s_1 \sqrt{3 a+s_1} \left(2 s_1-3 a\right)}\\
&+\frac{(N+2) \left(a-s_1\right) a_{\ell }^{\text{(2)}}\left(s_1\right)}{2 N s_1 \left(2 s_1-3 a\right)}\Bigg)~C^{(\a)}_{\ell}\left(\sqrt{\xi(s_1,a)}\right)\,,\\
\text{Disc}\mathcal{G}_{2}(s_1,s_2^{(+)}(s_1,a))&=\Phi(s_1;\a)\sum_{\ell=0}^{\infty}\left(2\ell+2\a\right)~3 \left(a-s_1\right) \Bigg(\frac{(N+2) a_{\ell }^{\text{(2)}}\left(s_1\right)-2 a_{\ell }^{\text{(0)}}\left(s_1\right)}{2 N s_1^2 \left(3 a-2 s_1\right)}\\
&+\frac{3 \sqrt{s_1-a} a_{\ell }^{\text{(1)}}\left(s_1\right)}{2 s_1^2 \left(3 a-2 s_1\right) \sqrt{3 a+s_1}}\Bigg)~C^{(\a)}_{\ell}\left(\sqrt{\xi(s_1,a)}\right)\,.
\end{split}
\ee
We use the same convention as \cite{ASAZ}. The $\a=\frac{d-3}{2}$, $\Phi(s_1;\a)=\Psi(\a)\frac{\sqrt{s_1+\frac{\mu}{3}}}{\left(s_1-\frac{2\mu}{3}\right)^{\a}}$ where $\Psi(\a)>0$ is  real and $\xi\left(s_{1}, a\right)=\xi_{0}+4 \xi_{0}\left(\frac{a}{s_{1}-a}\right),~~ \xi_0=\frac{s_1^2}{(s_1-2\mu/3)^2}$. The Gegenbauer polynomials can be expanded as $C_{\ell}^{(\a)}\left(\xi^{1 / 2}\right)=\sum_{j=0}^{\ell / 2} \frac{p_{\ell}^{(j)}\left(\xi_{0}\right)}{j !}\left(\xi-\xi_{0}\right)^{j}$ with $p_{\ell}^{(j)}\left(\xi_{0}\right)=\frac{\partial^j C^{(\a)}_{\ell}\left(\sqrt{\xi}\right)}{\partial{\xi^j}}{\bigg |}_{\xi=\xi_0}
$. Now we plug the formulas \eqref{eq:DiscG_k_aI} for $\text{Disc}\mathcal{G}_{k}$ in the equation \eqref{eq:G_k_ax}. After that we compute the the coefficient of $a^m$.  The coefficient of $a^m$ gives us formula
\be
\begin{split}\label{eq:cpqallform}
\mathcal{B}_{n,m,\ell}^{(0)}(s_1)=&\left(\frac{a_\ell^{(0)}(s_1)}{N}+\frac{(N-1)a_\ell^{(2)}(s_1)}{N}\right)~\frac{1}{\pi}\sum_{j=0}^{m}\frac{1}{s_1^{2 n+m}}\frac{p_{\ell}^{(j)}\left(\xi_{0}\right)}{j !}\left(4 \xi_{0}\right)^{j}\times\frac{(3 j-m-2n) (-n)_m}{(m-j)!(-n)_{j+1}}\,.
\end{split}
\ee
Similarly 
\be
\begin{split}\label{eq:cpqallform}
\mathcal{B}_{n,m,\ell}^{(1)}(s_1)=&\frac{1}{\pi}\sum_{j=0}^{m}\frac{1}{s_1^{2 n+m+1}}\frac{p_{\ell}^{(j)}\left(\xi_{0}\right)}{j !}\left(4 \xi_{0}\right)^{j}\times\frac{3}{2N}\Bigg[ (-1)^{m-j}\binom{n-j}{m-j} \left(2 a_{\ell }^{(0)}\left(s_1\right)-(N+2) a_{\ell }^{(2)}\left(s_1\right)\right)\\
&+N 3^{m-j} a_{\ell }^{(1)}\left(s_1\right) \Bigg\{\binom{-\frac{1}{2}}{m-j} \, _2F_1\left(j-m,j-n+\frac{1}{2};j-m+\frac{1}{2};-\frac{1}{3}\right)\\
&-\binom{-\frac{1}{2}}{-j+m-1} \, _2F_1\left(j-m+1,j-n+\frac{1}{2};j-m+\frac{3}{2};-\frac{1}{3}\right)\Bigg\}\Bigg]\,,
\end{split}
\ee
and
\be
\begin{split}\label{eq:cpqallform}
\mathcal{B}_{n,m,\ell}^{(2)}(s_1)=&\frac{1}{\pi}\sum_{j=0}^{m}\frac{1}{s_1^{2 n+m+2}}\frac{p_{\ell}^{(j)}\left(\xi_{0}\right)}{j !}\left(4 \xi_{0}\right)^{j}\times
\frac{3}{N} \Bigg[(-1)^{m-j} \binom{n-j}{m-j} \left((N+2) a_{\ell }^{(2)}\left(s_1\right)-2 a_{\ell }^{(0)}\left(s_1\right)\right)\\
&+N 3^{-j+m+1} \binom{-\frac{1}{2}}{m-j} a_{\ell }^{(1)}\left(s_1\right) \, _2F_1\left(j-m,j-n-\frac{1}{2};j-m+\frac{1}{2};-\frac{1}{3}\right)\Bigg]\,.
\end{split}
\ee
In the appendix, we have verified the dispersion relation and inversion formula against O(3) Lovelace-Shapiro model, see appendix \eqref{ap:LSsumrules}.

\subsubsection*{Sum rules}
In the equation \eqref{eq:GktoWk}, only positive powers of $x,y$ appears. Our dispersion relation is fully crossing symmetric by construction. After writing down the dispersion relations, there appear to be negative powers of $x$ in the expansion. Coefficients of such terms are $\mW^{(k)}_{n-m,m},~ m>n$. These terms $\mW^{(k)}_{n-m,m},~ m>n$ does not vanish on their own. Negative powers of  $x,y$, are non-local terms. Since there should not be any non-local terms, we should have
\be\label{sumrulev1}
\mW^{(k)}_{n-m,m}=0, \text{ for } m>n\,, k=0,1,2\,.
\ee
\textit{These conditions put very non-trivial constraints on the partial wave coefficients}. We call these sum rules \textit{locality constraints}. In the appendix, we have verified the sum rules against O(3) Lovelace-Shapiro model, see appendix \eqref{ap:LSsumrules}. 
\section{Positivity conditions on $\mathcal{W}_{p,q}^{(k)}$ }
In \cite{ASAZ} the positivity conditions on $\mathcal{W}_{p,q}^{(0)}$ were derived, using the positivity of the partial wave coefficients and the positivity of Gegenbauer polynomials,
\be\label{eq:positivityW0}
\begin{aligned}
&\sum_{r=0}^{m} \chi_{n,r, m}^{(0)}(\mu, \delta_0) \mathcal{W}_{n-r, r}^{(0)}  \geq 0\,, \\
&0 \leq \mathcal{W}_{n, 0}^{(0)}  \leq \frac{1}{\left(\delta_0+\frac{2 \mu}{3}\right)^{2}} \mathcal{W}_{n-1,0}^{(0)}\,, & n \geq 2\,.
\end{aligned}
\ee
Where $\d_0$ is the scale\footnote{For EFTs the analysis is modified by taking the lower limit of the integral $\frac{2\m}{3}\to\frac{2\m}{3}+\d_0$ , here $\d_0$ will be the EFT scale. Since for EFTs, most of the time the low energy amplitude $DiscG|_\frac{2\m}{3}^{\frac{2\m}{3}+\d_0}$ is computable. Therefore, we subtract the low energy part from the full  $DiscG$ and plug back in the integration , which changes the lower limit to $\frac{2\m}{3}+\d_0$. In our calculations, we will use $\d_0=0$. See \cite{ASAZ} for more discussions.} of the theory and 
\be
\begin{aligned}
&\chi_{n,m, m}^{(0)}(\mu, \delta) =1 \,, \chi_{n,r, m}^{(0)}(\mu, \delta) =\sum_{j=r+1}^{m}(-1)^{j+r+1} \chi_{n,j, m}^{(0)} \frac{\mathfrak{U}^{\alpha}{ }_{n, j, r}\left(\delta+\frac{2 \mu}{3}\right)}{\mathfrak{U}^{\alpha}{ }_{n, r, r}\left(\delta+\frac{2 \mu}{3}\right)}\,,\\
&\mathfrak{U}_{n, m, k}^{\alpha}=\sum_{k=0}^{m} \frac{\sqrt{16 \xi_{0}}^{k}(\alpha)_{k}(m+2 n-3 j) \Gamma(n-j) \Gamma(2 j-k)}{s_{1}^{m+2 n} \Gamma(k) j !(m-j) !(j-k) !(n-m) !}\,.
\end{aligned}
\ee 
There exist similar kinds of (but very non-trivial) positivity conditions for $\mathcal{W}_{p,q}^{(1)}$ and $\mathcal{W}_{p,q}^{(2)}$, which take the following form
\be \label{eq:positivityW1}
\sum_{r=0}^{m}\left(\Upsilon _{n,r,m}^{(1)}(\mu, \delta_0)\mathcal{W}_{n-r,r}^{(1)}+\chi_{n,r,m}^{(1)}(\mu, \delta_0)\mathcal{W}_{n-r,r}^{(0)}\right)\geq 0\,,
\ee
and
\be \label{eq:positivityW2}
\sum_{r=0}^{m}\left(\Upsilon _{n,r,m}^{(2)}(\mu, \delta_0)\mathcal{W}_{n-r,r}^{(2)}+\chi_{n,r,m}^{(2)}(\mu, \delta_0)\mathcal{W}_{n-r,r}^{(0)}\right)\geq 0\,,
\ee
with $\Upsilon _{n,m,m}^{(1)}(\mu, \delta_0)=1,~\Upsilon _{n,m,m}^{(2)}(\mu, \delta_0)=1$.

The $\chi_{n,r,m}^{(1)}(\mu, \delta_0),~\chi_{n,r,m}^{(2)}(\mu, \delta_0),~\Upsilon _{n,r,m}^{(1)}(\mu, \delta_0),~\Upsilon _{n,r,m}^{(2)}(\mu, \delta_0)$ are known positive coefficients. Its quite hard to get a general formula, but one can work out case by case in $m$. For example, below, we listed them up to $m=3$.
\be
\begin{array}{|l|l|l|}
\hline
 \chi _{n,0,1}^{\text{(1)}}(\delta ,\mu )=\frac{27 (2 n+9) (N+2)}{8 (N-1) (3 \delta +2 \mu )^2} & \Upsilon _{n,0,1}^{\text{(1)}}(\delta ,\mu )=\frac{3 (n+4)}{3 \delta +2 \mu } & \chi _{n,1,1}^{\text{(1)}}(\delta ,\mu )=\frac{9 (N+2)}{4 (N-1) (3 \delta +2 \mu )} \\
\hline
 \chi _{n,0,2}^{\text{(1)}}(\delta ,\mu )=\frac{81 (2 n (n+10)+51) (N+2)}{16 (N-1) (3 \delta +2 \mu )^3} & \chi _{n,1,2}^{\text{(1)}}(\delta ,\mu )=\frac{27 (2 n+9) (N+2)}{8 (N-1) (3 \delta +2 \mu )^2} & \Upsilon _{n,0,2}^{\text{(1)}}(\delta ,\mu )=\frac{9 (n+4) (n+5)}{2 (3 \delta +2 \mu )^2} \\
\hline
 \Upsilon _{n,1,2}^{\text{(1)}}(\delta ,\mu )=\frac{3 (n+4)}{3 \delta +2 \mu } & \chi _{n,2,2}^{\text{(1)}}(\delta ,\mu )=\frac{9 (N+2)}{4 (N-1) (3 \delta +2 \mu )} & \chi _{n,0,3}^{\text{(1)}}(\delta ,\mu )=\frac{81 (2 n (n (2 n+33)+172)+879) (N+2)}{32 (N-1) (3 \delta +2 \mu )^4} \\
\hline
 \chi _{n,1,3}^{\text{(1)}}(\delta ,\mu )=\frac{81 (2 n (n+10)+47) (N+2)}{16 (N-1) (3 \delta +2 \mu )^3} & \chi _{n,2,3}^{\text{(1)}}(\delta ,\mu )=\frac{27 (2 n+9) (N+2)}{8 (N-1) (3 \delta +2 \mu )^2} & \Upsilon _{n,0,3}^{\text{(1)}}(\delta ,\mu )=\frac{9 (n (n (n+15)+68)+168)}{2 (3 \delta +2 \mu )^3} \\
\hline
 \Upsilon _{n,1,3}^{\text{(1)}}(\delta ,\mu )=\frac{9 (n+3) (n+6)}{2 (3 \delta +2 \mu )^2} & \Upsilon _{n,2,3}^{\text{(1)}}(\delta ,\mu )=\frac{3 (n+4)}{3 \delta +2 \mu } & \chi _{n,3,3}^{\text{(1)}}(\delta ,\mu )=\frac{9 (N+2)}{4 (N-1) (3 \delta +2 \mu )} \\
\hline
\end{array}
\ee
and

\be
\begin{array}{|l|l|l|}
\hline
 \chi _{n,0,1}^{\text{(2)}}(\delta ,\mu )=\frac{81 (2 n+5)}{4 (3 \delta +2 \mu )^3} & \Upsilon _{n,0,1}^{\text{(2)}}=\frac{3 (n+2)}{3 \delta +2 \mu } & \chi _{n,1,1}^{\text{(2)}}(\delta ,\mu )=\frac{27}{2 (3 \delta +2 \mu )^2} \\
\hline
 \chi _{n,0,2}^{\text{(2)}}(\delta ,\mu )=\frac{243 (2 n (n+6)+7)}{8 (3 \delta +2 \mu )^4} & \chi _{n,1,2}^{\text{(2)}}(\delta ,\mu )=\frac{81 (2 n+5)}{4 (3 \delta +2 \mu )^3} & \Upsilon _{n,0,2}^{\text{(2)}}(\delta ,\mu )=\frac{9 n (n+5)}{2 (3 \delta +2 \mu )^2} \\
\hline
 \Upsilon _{n,1,2}^{\text{(2)}}(\delta ,\mu )=\frac{3 (n+2)}{3 \delta +2 \mu } & \chi _{n,2,2}^{\text{(2)}}(\delta ,\mu )=\frac{27}{2 (3 \delta +2 \mu )^2} & \chi _{n,0,3}^{\text{(2)}}(\delta ,\mu )=\frac{243 (2 n (n (2 n+21)+28)+51)}{16 (3 \delta +2 \mu )^5} \\
\hline
 \chi _{n,1,3}^{\text{(2)}}(\delta ,\mu )=\frac{243 (2 n (n+6)+3)}{8 (3 \delta +2 \mu )^4} & \chi _{n,2,3}^{\text{(2)}}(\delta ,\mu )=\frac{81 (2 n+5)}{4 (3 \delta +2 \mu )^3} & \Upsilon _{n,0,3}^{\text{(2)}}(\delta ,\mu )=\frac{9 n (n (n+9)+2)}{2 (3 \delta +2 \mu )^3} \\
\hline
 \Upsilon _{n,1,3}^{\text{(2)}}(\delta ,\mu )=\frac{9 (n (n+5)-2)}{2 (3 \delta +2 \mu )^2} & \Upsilon _{n,2,3}^{\text{(2)}}(\delta ,\mu )=\frac{3 (n+2)}{3 \delta +2 \mu } & \chi _{n,3,3}^{\text{(2)}}(\delta ,\mu )=\frac{27}{2 (3 \delta +2 \mu )^2} \\
\hline
\end{array}
\ee
These positivity conditions follow from the positivity of partial wave coefficients $a^{(I)}_\ell(s_1)$ presented in the equation \eqref{eq:positiveaell} and the positivity of Gegenbauer polynomials. Proof and derivation of these positivity conditions can be found in appendix \eqref{ap:positivity}.
\section{Geometric function theory for $O(N)$ model}
\subsection{Typically real functions}
A typically real function $f(z)$ satisfies 
$\text{Im}[f(z)] \text{Im}[z] > 0, ~~ \text{Im}[z]\neq 0$, \textit{i.e.}  the  imaginary part of the function is positive in the upper half plane, and negative in the lower half plane. We will consider the subclass of typically real functions, namely regular and typically real inside the unit disk $|z|<1$
\be
f(z)=z+\sum_{p=2}^{\infty}b_p z^p\,,
\ee
A schlicht function $f(z)$ (normalized $f(0)=0$ and $f'(0)=1$, see \cite{PHASAZ} for a quick overview), which is univalent inside the unit disk $|z|<1$ with $b_p\in \mathbb{R}$, is also typically real function. The kernel $H$ is a function of this kind.

If $f(z)$ is a regular typically real function in $|z|<1$, then the coefficients should satisfy the bounds (see \cite{PRAS, komatu} for proof)
\be\label{eq:Rogbounds}
-\kappa_p \leq b_p \leq p \,.
\ee
An important representation of typically real is the Robertson representation \cite{Robertson}.
\subsubsection*{Robertson representation:}

\textit{In $|z| < 1$ a regular function $F(z)$ is typically
real if and only if it has the Robertson representation:
\be
F(z)=\int_{-1}^{1} d f(\eta) \frac{z}{1-2 \eta z+z^{2}}\,,
\ee
where the measure $f(\eta)$ is a non-decreasing function.}

From the crossing symmetric dispersion relation, we get the full amplitude as an integral of discontinuity of the amplitude $\mathcal{A}(s_1,s^{(+)}(s_1,a))$ times the kernel $H(s_1';s_2,s_3)$. The kernel $H(s_1';s_2,s_3)$ is a univalent typically real function. For some range of $a$, the $\mathcal{A}(s_1,s^{(+)}(s_1,a))$ is non-negative. We will show below from these two facts that the full amplitude can be recast as Robertson representation; hence it is a typically real function.

\subsection{Typically real functions for $O(N)$ model}

We would consider the combinations 
\be\label{eq:FktoGk}
\begin{split}
&\mathcal{F}_0(\tilde{z},a)=\mathcal{G}_0(s_1,s_2)-\a_0^{(0)}\,,\\
&\mathcal{F}_1(\tilde{z},a)= \mathcal{G}_1(s_1,s_2)+\mathcal{G}_0(s_1,s_2)-\a_0^{(1)}-\a_0^{(0)}\,, \\
&\mathcal{F}_2(\tilde{z},a)=\mathcal{G}_2(s_1,s_2)+\frac{\mathcal{G}_0(s_1,s_2)}{3}-\a_0^{(2)}-\frac{\a_0^{(0)}}{3}\,.
\end{split}
\ee
By definition then we can write ($\zt=z^3$)
\be
\mathcal{F}_k(\zt,a)=\sum_{n=1}^{\infty}\a_n^{(k)}(a)a^{2n}\zt^n\,;\,\, k=0,1,2\,,
\ee
with
\be\label{eq:alphaktoWk}
\begin{split}
&\alpha_{p}^{(k)}(a) a^{2 p}=\sum_{n=0}^{p} \sum_{m=0}^{n}\widehat{\mathcal{W}}_{n-m,m}^{(k)}a^{m}(-1)^{p-n}(-27)^{n} a^{2 n}\left(\begin{array}{c}
-2 n \\
p-n
\end{array}\right)\,,\\
&\widehat{\mathcal{W}}_{n-m,m}^{(k)}=\left(1-\frac{2\d_{2,k}}{3}\right)\mathcal{W}_{n-m, m}^{(0)}+\d_{1,k}\mathcal{W}_{n-m, m}^{(1)} +\d_{2,k}\mathcal{W}_{n-m, m}^{(2)}\,.
\end{split}
\ee
In \cite{PRAS} it was shown that $\mathcal{F}_0(\tilde{z},a)$ is a typically real function from the positivity of the $\immpGo\,.$
\\

The kernel can be written as in $\zt$ variables as 
\be
H\left(s_{1},s_1,s_2,s_3\right)=\frac{27 a^{2} \tilde{z}\left(2 s_{1}-3 a\right)}{27 a^{3} \tilde{z}-27 a^{2} \tilde{z} s_{1}-(\tilde{z}-1)^{2} s_{1}^{3}}=\sum_{n=0}^{\infty} \beta_{n}\left(a, s_{1}\right) \tilde{z}^{n}\,.
\ee
We know from \cite{PHASAZ} that Kernel is a univalent function inside the unit disk. 
 Notice that for  $\beta_{1}\left(a, s_{1}\right)=\frac{27 a^{2}}{s_{1}^{3}}\left(3 a-2 s_{1}\right)<0$, we must have $a<4\m/9$, since $s_1\geq 2\m/3$ (see \cite{PRAS}). The dispersion relations for $\mathcal{F}_k(\tilde{z},a)$ are given by
\be
\begin{split}
&\mathcal{F}_{k}(\zt,a)=\frac{1}{\pi} \int_{\frac{2\m}{3} }^{\infty} \frac{d s_1^{\prime}}{s_1^{\prime}} \text{Disc}\left[ \left(1-\frac{2 \delta _{2,k}}{3}\right)\mathcal{G}_0+\delta _{1,k}\mathcal{G}_1 +\delta _{2,k}\mathcal{G}_2 \right]\left(s_1^{\prime} ; s_2^{(+)}\left(s_1^{\prime} ,a\right)\right) \frac{27 a^{2} \tilde{z}\left(2 s_{1}'-3 a\right)}{27 a^{3} \tilde{z}-27 a^{2} \tilde{z} s_{1}'-(\tilde{z}-1)^{2} s_{1}^{' 3}}\,.
\end{split}
\ee
In the combination $\text{Disc}\left[ \left(1-\frac{2 \delta _{2,k}}{3}\right)\mathcal{G}_0+\delta _{1,k}\mathcal{G}_1 +\delta _{2,k}\mathcal{G}_2 \right]$, if we use \eqref{eq:DiscG_k} and collect the coefficients of $\mathcal{A}^{(I)}(s_1,s^{(+)}(s_1,a))$, we find that the coefficients are always positive for $~s_1'\geq \frac{2\mu}{3},~ N\geq 3,~\mu\geq 4$ for some range of $a$: for $k=1,2$ the range of $a$  is $-\frac{s_1'}{3}<a<\frac{s_1'}{3}$, while for $k=0$ the range of $a$ is $-\frac{s_1'}{3}<a<\frac{2s_1'}{3}$. Now since of the absorptive part of isospin amplitudes is always non-negative
$
\mathcal{A}^{(I)}\left(s_1^{\prime} ; s_2^{(+)}\left(s_1^{\prime} ,a\right)\right)
$ for $-\frac{2\m}{9}<a<\frac{2\mu}{3},~s_1'\geq \frac{2\mu}{3}$.  Therefore we find that the combination is always non-negative for ranges discussed above
\be
\begin{split}
\text{Disc}\left[ \left(1-\frac{2 \delta _{2,k}}{3}\right)\mathcal{G}_0+\delta _{1,k}\mathcal{G}_1 +\delta _{2,k}\mathcal{G}_2 \right]\left(s_1^{\prime} ; s_2^{(+)}\left(s_1^{\prime} ,a\right)\right)\geq 0\,.
\end{split}
\ee
Since $s_1'\geq \frac{2\mu}{3}$, therefore the range of $a$ in case of $k=0$ is given by $-\frac{2\m}{9}<a<\frac{4\mu}{9}$, while in case of $k=1,2$ range of $a$ is given by $-\frac{2\m}{9}<a<\frac{2\mu}{9}$. See appendix \eqref{ap:positivityofF} for more clarifications.

We change the variable $s_1'$ to $\eta$ by the equation
\be
-\frac{27 a^{3}}{\left(s_{1}^{\prime}\right)^{3}}+\frac{27 a^{2}}{\left(s_{1}^{\prime}\right)^{2}}-2=2 \eta\,.
\ee
In this changed variable, we can write $\mathcal{F}_k(\tilde{z},a)$ as (see \cite{PRAS} for more details)
\be 
\frac{\mathcal{F}_{k}(\tilde{z}, a)}{\int^{1}_{-1}d\eta ~\overline{\text{Disc}\mathcal{F}}_k(\eta,a)}= \int_{-1}^{1} d f_k(\eta) ~~\frac{\tilde{z}}{1-2 \eta \tilde{z}+\tilde{z}^{2}}\,,
\ee
where
\be
d f_k(\eta)=\frac{\overline{\text{Disc}\mathcal{F}}_k(\eta,a)~d \eta}{\int^{1}_{-1}d\eta ~\overline{\text{Disc}\mathcal{F}}_k(\eta,a)}\,.
\ee
We have adopted the notation $\overline{\text{Disc}\mathcal{F}}_k(\eta,a)$ for  $\text{Disc}\left[ \left(1-\frac{2 \delta _{2,k}}{3}\right)\mathcal{G}_0+\delta _{1,k}\mathcal{G}_1 +\delta _{2,k}\mathcal{G}_2 \right]\left(s_1^{\prime} ; s_2^{(+)}\left(s_1^{\prime} ,a\right)\right)\,,$ after changing the variable $s_1'\to \eta$. 
Since the $\overline{\text{Disc}\mathcal{F}}_k(\eta,a)$ are positive (for $s_1'\geq \frac{2\mu}{3},~ N\geq 3,~\mu\geq 4$ and range of $a$ in case of $k=0$ is given by $-\frac{2\m}{9}<a<\frac{4\mu}{9}$, while in case of $k=1,2$ range of $a$ is given by $-\frac{2\m}{9}<a<\frac{2\mu}{9}$) then 
\be\label{eq:RobRepfinal}
 f_k(\zeta)=\frac{\int^{\zeta}_{-1}\overline{\text{Disc}\mathcal{F}}_k(\eta,a)~d \eta}{\int^{1}_{-1}d\eta ~\overline{\text{Disc}\mathcal{F}}_k(\eta,a)}\,.
\ee
are non-decreasing functions. We can conclude that functions $\mathcal{F}_k(\zt,a)$ are typically real functions. 

\subsection{Rogosinski bounds on $\a_p^{(k)}(a)$}

Since functions $\mathcal{F}_k(\zt,a)$ are typically real functions, therefore we can readily write
\be\label{eq:Rogbounds}
\begin{split}
&-\kappa_n \leq \frac{\a_n^{(k)}(a)a^{2n}}{\a_1^{(k)}(a)a^2}\leq n \,,\text{  for   } \begin{cases}
& k=0\,, -\frac{2\m}{9}<a<\frac{4\m}{9}\,,\\
& k=1,2\,, -\frac{2\m}{9}<a<\frac{2\m}{9}\,,\\
\end{cases}\,,\\
\end{split}
\ee
where
\be 
\kappa_{p}=-\frac{\sin p \varphi_{p}}{\sin \varphi_{p}}, \quad \,\,\csc \left(\varphi _p\right) \left(p \cos \left(p \varphi _p\right)-\sin \left(p \varphi _p\right) \cot \left(\varphi _p\right)\right)=0 \,.
\ee
For  $p$ odd, $\pi/p<\varphi_{p}<3\pi/2p$, while for $p$ even, $\varphi_{p}=\pi$ is only solution, giving $\kappa_p = p$. Notice the range of $a$ is different for different $k$ and for $k=1,2$, one will have to restrict to the cases $N\geq 3\,,\m\geq 4$
\section{Geometric function theory constraints on Pion amplitudes}
In this section, we discuss the geometric function theory constraints on the Pion amplitudes. We will consider the case $N=3, \mu=4$ in this section. For pion amplitudes, the functions $\mathcal{F}_k(\zt,a)$ are indeed typically real functions. It is more natural to work with $\widehat{\mathcal{W}}_{n-m,m}^{(k)}$ defined in \eqref{eq:alphaktoWk}.

\subsection{Bounds on coefficients $\widehat{\mathcal{W}}_{n-m,m}^{(k)}$}
Since the imaginary part of $\mathcal{F}_k(\zt,a)$ is positive for the range given in \eqref{eq:Rogbounds}, therefore we note that
\be
\wmW^{(k)}_{n,0}\geq 0 \text{  for  } \m\geq 4\,.
\ee
This is true since here $m=0$ or we are considering the coefficients of $a^0$. Which enable us to compute $\wmW^{(k)}_{n,0}$ , which is given by
\be
\wmW^{(k)}_{n,0}=\frac{1}{\pi} \int_{\frac{2\m}{3} }^{\infty} \frac{d s_1^{\prime}}{s_1^{\prime 2n+1}} \left(2\times \text{Disc}\left[ \left(1-\frac{2 \delta _{2,k}}{3}\right)\mathcal{G}_0+\delta _{1,k}\mathcal{G}_1 +\delta _{2,k}\mathcal{G}_2 \right]\left(s_1^{\prime} ; 0\right)\right)\geq 0\,.
\ee
Therefore we can write (derivation is similar to second equation of \eqref{eq:positivityW0}, see \cite{ASAZ} for details)
\be
0\leq \wmW^{(k)}_{n,0}\leq \frac{\wmW^{(k)}_{n-1,0}}{(\d_0+\frac{2\m}{3})^2}\,.
\ee
All the bounds come in the form of  $\frac{\wmW^{(k)}_{p,q}}{\wmW^{(k)}_{1,0}}$ (we can divide since $\wmW^{(k)}_{1,0}>0$). Therefore we will normalize $\wmW^{(k)}_{p,q}$ such that
\be 
\wmW^{(k)}_{1,0}=1\,.
\ee

We can use the Rogosinski bounds on $\a_p^{(k)}(a)a^{2p}$ and positivity bounds on the $\mathcal{W}^{(k)}_{p,q}$, to put two sided bounds on $\widehat{\mathcal{W}}^{(k)}_{p,q}$. Table \eqref{tab:boundsWk} shows first few bounds on $\widehat{\mathcal{W}}^{(k)}_{p,q}$. For clarity, we write all the equations once again here.
\\
\textbf{Rogosinski bounds on $\a_p^{(k)}(a)a^{2p}$:}
\be\label{eq:RogoAllk}
\begin{split}
&-\kappa_n \leq \frac{\a_n^{(k)}(a)a^{2n}}{\a_1^{(k)}(a)a^2}\leq n \,,\text{  for   } \begin{cases}
& k=0\,, -\frac{2\m}{9}<a<\frac{4\m}{9}\,,\\
& k=1,2\,, -\frac{2\m}{9}<a<\frac{2\m}{9}\,,\\
\end{cases}\,,\\
&\alpha_{p}^{(k)}(a) a^{2 p}=\sum_{n=0}^{p} \sum_{m=0}^{n}\widehat{\mathcal{W}}_{n-m,m}^{(k)}a^{m}(-1)^{p-n}(-27)^{n} a^{2 n}\left(\begin{array}{c}
-2 n \\
p-n
\end{array}\right)\,,\\
&\widehat{\mathcal{W}}_{n-m,m}^{(k)}=\left(1-\frac{2\d_{2,k}}{3}\right)\mathcal{W}_{n-m, m}^{(0)}+\d_{1,k}\mathcal{W}_{n-m, m}^{(1)} +\d_{2,k}\mathcal{W}_{n-m, m}^{(2)}\,.
\end{split}
\ee
\\
\textbf{Positivity conditions:}
\be\label{eq:Positivityallk}
\begin{aligned}
&\sum_{r=0}^{m} \chi_{n,r, m}^{(0)}(\mu, \delta_0) \mathcal{W}_{n-r, r}^{(0)}  \geq 0 \,,~~~\sum_{r=0}^{m}\left(\Upsilon _{n,r,m}^{(1)}(\mu, \delta_0)\mathcal{W}_{n-r,r}^{(1)}+\chi_{n,r,m}^{(1)}(\mu, \delta_0)\mathcal{W}_{n-r,r}^{(0)}\right)\geq 0\,,\\
&\sum_{r=0}^{m}\left(\Upsilon _{n,r,m}^{(2)}(\mu, \delta_0)\mathcal{W}_{n-r,r}^{(2)}+\chi_{n,r,m}^{(2)}(\mu, \delta_0)\mathcal{W}_{n-r,r}^{(0)}\right)\geq 0\,,~~~0\leq \wmW^{(k)}_{n,0}\leq \frac{\wmW^{(k)}_{n-1,0}}{(\d_0+\frac{2\m}{3})^2}\,.
\end{aligned}
\ee
In the above equations, we will consider $m=1,2,3,\dots n$. The Taylor coefficients of the amplitude around the crossing symmetric point (appearing on $\a_n^{(k)}(a)a^{2n}$) can be bounded by using up to $n$-th equations in Rogosinski bounds on $\a_n^{(k)}(a)a^{2n}$; in that case, one must use up to $n$-th positivity conditions with $m=1,2,3,\dots n$.

\subsubsection*{The $n=2$ bounds}
The simplest exercise is for the $n=2$ bounds. We put $n=2$ in equation  \eqref{eq:RogoAllk} and $n=2,m=1,2$ in equation \eqref{eq:Positivityallk}, then maximize and minimize\footnote{In Mathematica, we define ImplicitRegion, and we find the bounds on the Taylor coefficients using RegionBounds.}
 with respect to the coefficients $\widehat{\mathcal{W}}_{n-m,m}^{(k)}$ that appear in those equations. 
Table \eqref{tab:boundsWkn2} contains the bounds we have found.

\begin{table}[hbt!]
\centering
\begin{tabular}{|L|L|L|L|L|L|L|L|L|}
\hline
\widehat{\mathcal{W}}_{p,q}^{(0)} & \text{Lower} & \text{Upper} & \widehat{\mathcal{W}}_{p,q}^{(1)} & \text{Lower} & \text{Upper} & \widehat{\mathcal{W}}_{p,q}^{(2)} & \text{Lower} & \text{Upper} \\
\hline
 \widehat{\mathcal{W}}_{1,1}^{\text{(0)}} & -0.131836 & 0.105469 & \widehat{\mathcal{W}}_{1,1}^{\text{(1)}} & -0.211282 & 0.217698 & \widehat{\mathcal{W}}_{1,1}^{\text{(2)}} & -0.147396 & 0.217698 \\
\hline
 \widehat{\mathcal{W}}_{2,0}^{\text{(0)}} & 0 & 0.140625 & \widehat{\mathcal{W}}_{2,0}^{\text{(1)}} & 0 & 0.140625 & \widehat{\mathcal{W}}_{2,0}^{\text{(2)}} & -0.0000601055 & 0.140625 \\
\hline
 \widehat{\mathcal{W}}_{0,2}^{\text{(0)}} & -0.0889893 & 0.0395508 & \widehat{\mathcal{W}}_{0,2}^{\text{(1)}} & -0.180231 & 0.243389 & \widehat{\mathcal{W}}_{0,2}^{\text{(2)}} & -0.165266 & 0.243427 \\
\hline
 \widehat{\mathcal{W}}_{0,1}^{\text{(0)}} & -0.562500 & 1.12500 & \widehat{\mathcal{W}}_{0,1}^{\text{(1)}} & -1.13924 & 1.12500 & \widehat{\mathcal{W}}_{0,1}^{\text{(2)}} & -1.09570 & 1.12500 \\
\hline
\end{tabular}
\caption{Bounds on $\widehat{\mathcal{W}}^{(k)}_{p,q}$ for $n=2$ in equation \eqref{eq:Rogbounds}. We use the normalizations $\widehat{\mathcal{W}}^{(k)}_{1,0}=1,$. The normalization can be restored via replacing $\widehat{\mathcal{W}}_{p,q}^{(k)}\to\frac{\widehat{\mathcal{W}}_{p,q}^{(k)}}{\widehat{\mathcal{W}}_{1,0}^{(k)}}$.}
\label{tab:boundsWkn2}
\end{table}

\begin{figure}[hbt!]
\centering
\includegraphics[scale=0.5]{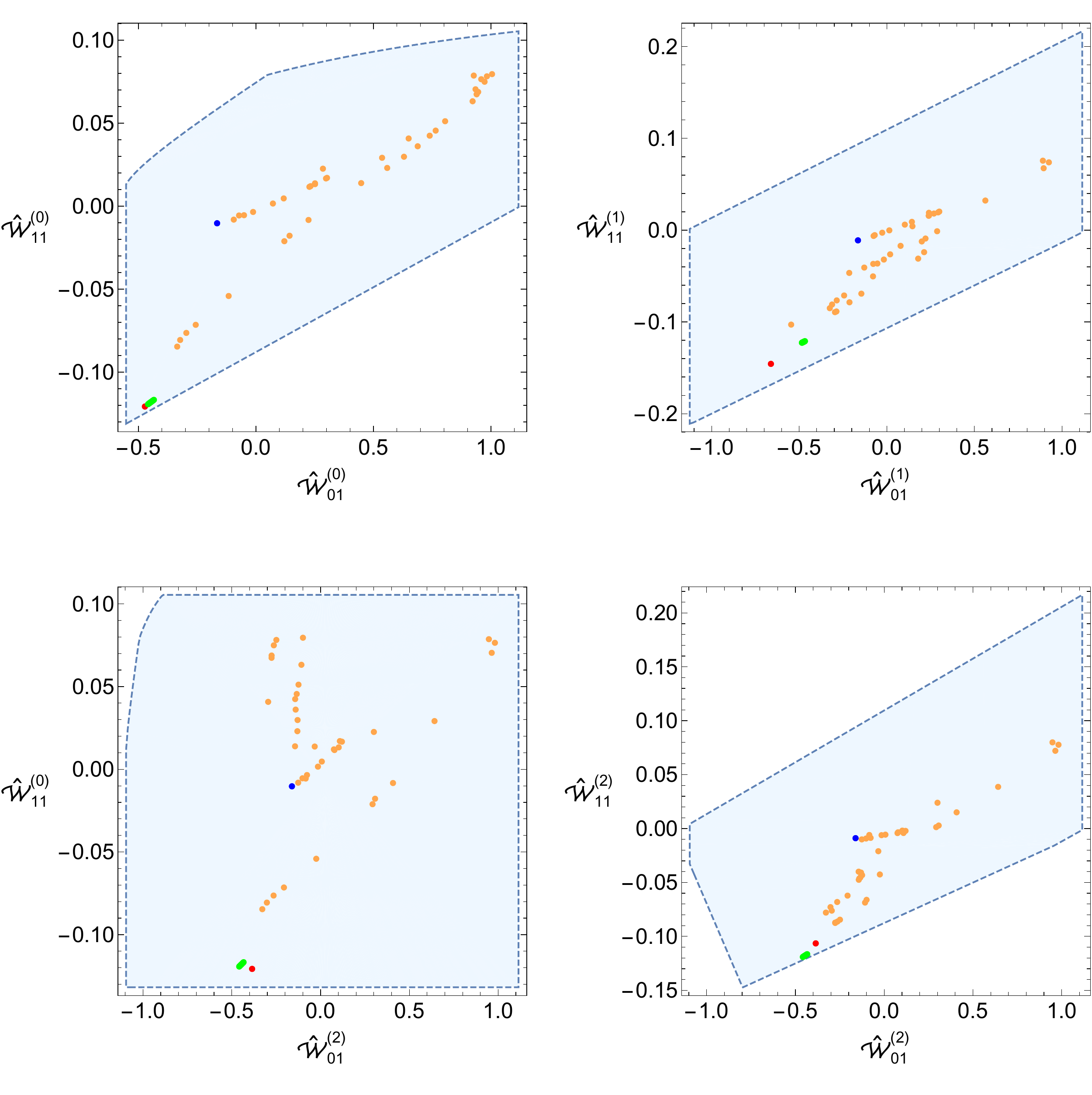}
\caption{The light blue region is the theory space allowed by positivity and typically real-ness. Some of the known theories has been indicated. Green region is the $O(3)$ Lovelace-Shapiro model \eqref{LS_model_full} with $\b'=\frac{1}{m^2_\rho}(1-\b_0)$ and $\beta_0$ varies from $0.465<\b_0<0.489$ (see \cite{ABAS}). The red dot is the the $O(3)$ Lovelace-Shapiro model \eqref{LS_model_full} with $\b_0=\b'=1/2$. The blue dot is the 2-loop chiral perturbation theory with parameters are taken from experimental values values. The orange regions are obtained from the S-matrix bootstrap   amplitude for the upper river boundaries \cite{ABPHAS,ABAS}}
\label{fig:finiteregion}
\end{figure}

Using the $n=2$ constraints, we find the finite region of theory space as depicted in figure \eqref{fig:finiteregion}. We have also indicated some of the known theories.

\subsubsection*{The $n=3$ bounds}
We put $n=2,3$ in equation \eqref{eq:RogoAllk} and $n=2,3,m=1,2,3$ in equation \eqref{eq:Positivityallk}, then Maximize and Minimize with respect to the coefficients $\widehat{\mathcal{W}}_{n-m,m}^{(k)}$ those appear in these equations, we find the bounds given in \eqref{tab:boundsWk}.
\begin{table}[hbt!]
\centering
\begin{tabular}{|L|L|L||L|L|L||L|L|L|}
\hline
 \widehat{\mathcal{W}}_{p,q}^{(0)} & \text{Lower} & \text{Upper} & \widehat{\mathcal{W}}_{p,q}^{(1)} & \text{Lower} & \text{Upper} & \widehat{\mathcal{W}}_{p,q}^{(2)} & \text{Lower} & \text{Upper} \\
\hline
 \widehat{\mathcal{W}}_{1,1}^{\text{(0)}} & -0.131836 & 0.105469 & \widehat{\mathcal{W}}_{1,1}^{\text{(1)}} & -0.211282 & 0.217698 & \widehat{\mathcal{W}}_{1,1}^{\text{(2)}} & -0.147396 & 0.217698 \\
\hline
 \widehat{\mathcal{W}}_{2,0}^{\text{(0)}} & 0 & 0.140625 & \widehat{\mathcal{W}}_{2,0}^{\text{(1)}} & 0 & 0.140625 & \widehat{\mathcal{W}}_{2,0}^{\text{(2)}} & 0 & 0.140625 \\
\hline
 \widehat{\mathcal{W}}_{0,2}^{\text{(0)}} & -0.0889893 & 0.0395508 & \widehat{\mathcal{W}}_{0,2}^{\text{(1)}} & -0.180231 & 0.243389 & \widehat{\mathcal{W}}_{0,2}^{\text{(2)}} & -0.165266 & 0.243427 \\
\hline
 \widehat{\mathcal{W}}_{0,1}^{\text{(0)}} & -0.562500 & 1.12500 & \widehat{\mathcal{W}}_{0,1}^{\text{(1)}} & -1.13924 & 1.12500 & \widehat{\mathcal{W}}_{0,1}^{\text{(2)}} & -1.09570 & 1.12500 \\
\hline
 \widehat{\mathcal{W}}_{1,2}^{\text{(0)}} & -0.0194715 & 0.0114256 & \widehat{\mathcal{W}}_{1,2}^{\text{(1)}} & -0.0372450 & 0.0475034 & \widehat{\mathcal{W}}_{1,2}^{\text{(2)}} & -0.0291138 & 0.0648537 \\
\hline
 \widehat{\mathcal{W}}_{2,1}^{\text{(0)}} & -0.0259552 & 0.0104932 & \widehat{\mathcal{W}}_{2,1}^{\text{(1)}} & -0.0371815 & 0.100608 & \widehat{\mathcal{W}}_{2,1}^{\text{(2)}} & -0.0290552 & 0.100608 \\
\hline
 \widehat{\mathcal{W}}_{3,0}^{\text{(0)}} & 0 & 0.0197754 & \widehat{\mathcal{W}}_{3,0}^{\text{(1)}} & 0 & 0.0197754 & \widehat{\mathcal{W}}_{3,0}^{\text{(2)}} & -0.0423541 & 0.0197754 \\
\hline
 \widehat{\mathcal{W}}_{0,3}^{\text{(0)}} & -0.00587333 & 0.00708816 & \widehat{\mathcal{W}}_{0,3}^{\text{(1)}} & -0.110787 & 0.0599811 & \widehat{\mathcal{W}}_{0,3}^{\text{(2)}} & -0.110787 & 0.0546385 \\
\hline
\end{tabular}
\caption{Bounds on $\widehat{\mathcal{W}}^{(k)}_{p,q}$ for $n=3$.}
\label{tab:boundsWk}
\end{table}

Once we have bounds on $\widehat{\mathcal{W}}^{(k)}_{p,q}\,$, in principle we can bound Taylor coefficients of the amplitude around the crossing symmetric point of any physical reactions. Below we show such an example.

Even though we work with $N=3$, our bounds (table \eqref{tab:boundsWkn2}, \eqref{tab:boundsWk}) do not depend on $N$. The $N$ dependant factor only enters through the positivity condition \eqref{eq:positivityW1}, which goes away when considering the ratio.

\subsection{Bounds on Taylor coefficients of physical amplitudes}
The $A(s_1 \mid s_2, s_3) $ is symmetric under exchange of $s_2\leftrightarrow s_3$. Therefore without lose of generality, we can write
\be
A(s_1 \mid s_2, s_3) =\sum_{p=0,q=0}^{\infty}\mathcal{C}_{p,q}(-s_2 s_3)^p (-s_1)^q\,.
\ee

\begin{figure}[hbt!]
\centering
\includegraphics[scale=0.6]{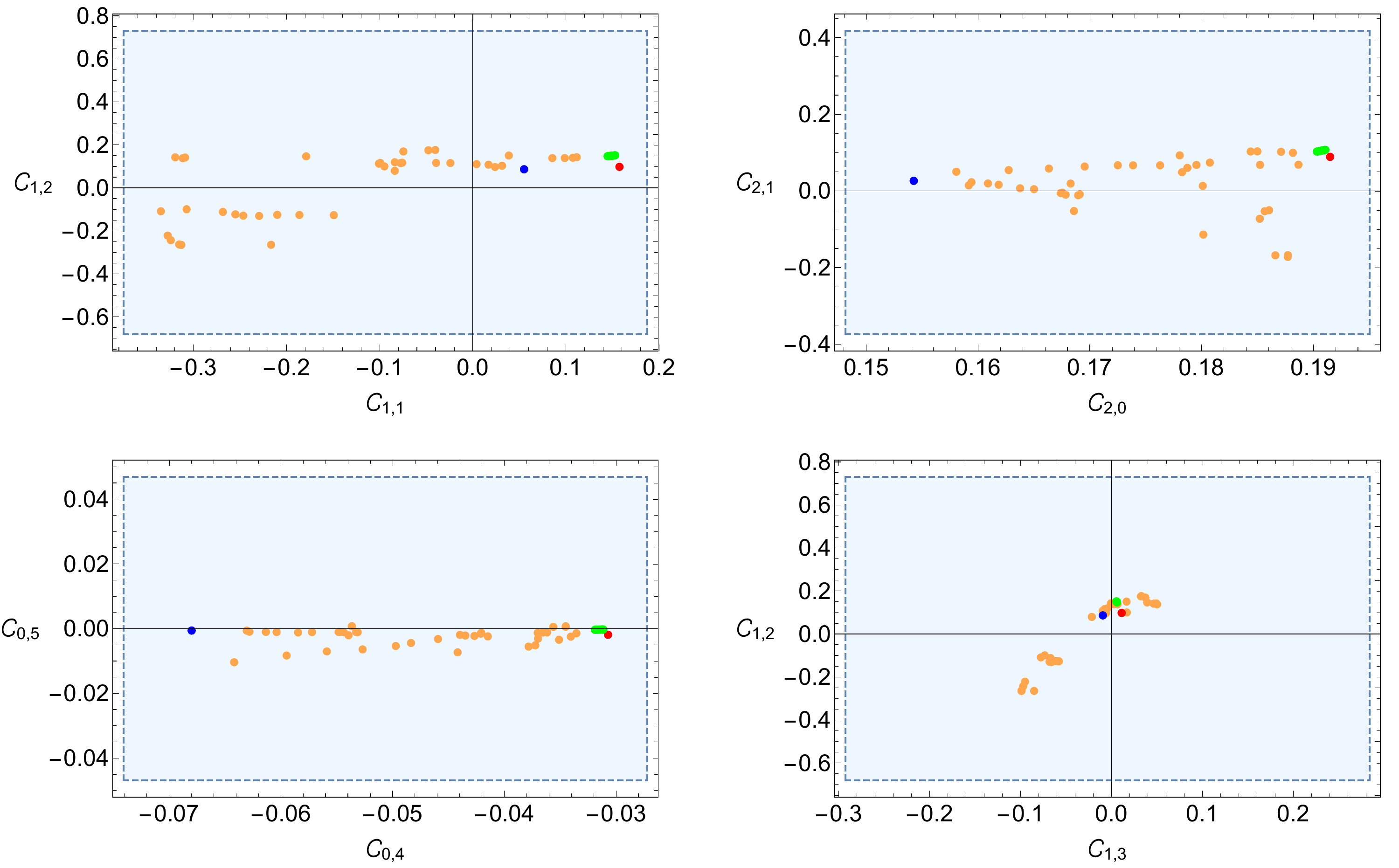}
\caption{The light blue region is the theory space allowed by the constraints in the table \eqref{tab:Cpqbounds}. Some of the known theories are indicated. Green region is the $O(3)$ Lovelace-Shapiro model \eqref{LS_model_full} with $\b'=\frac{1}{m^2_\rho}(1-\b_0)$ and $\beta_0$ varies from $0.465<\b_0<0.489$ (see \cite{ABAS}). The red dot is the the $O(3)$ Lovelace-Shapiro model \eqref{LS_model_full} with $\b_0=\b'=1/2$. The blue dot is the 2-loop chiral perturbation theory with parameters are taken from experimental values values. The orange regions are obtained from the S-matrix bootstrap  amplitude for the upper river boundaries \cite{ABPHAS,ABAS}.}
\label{fig:finiteregionCs}
\end{figure}

\textit{Once we have the bounds on $\widehat{\mathcal{W}}^{(k)}_{p,q}$, using the isospin amplitudes, namely using \eqref{eq:isotoGk}, we can put bounds on Taylor coefficients of any physical amplitudes around the crossing symmetric point.} For example, for the reaction $\pi^{+}+\pi^{-} \rightarrow \pi^{0}+\pi^{0}$ the amplitude is
\be
\mathcal{M}^{({\tiny \pi^{+}+\pi^{-} \rightarrow \pi^{0}+\pi^{0}})}(s_1,s_2)=\frac{1}{3}\left(\mathcal{M}^{0}(s_1,s_2)-\mathcal{M}^{2}(s_1,s_2)\right)=A(s_1 \mid s_2, s_3)=\sum_{p=0,q=0}^{\infty}\mathcal{C}_{p,q}(-s_2 s_3)^p (-s_1 )^q\,.
\ee
Using the equation \eqref{eq:isotoGk} and the expansion \eqref{eq:GktoWk}, we find that
\be
\frac{1}{3}\left(\mathcal{M}^{0}(s_1,s_2)-\mathcal{M}^{2}(s_1,s_2)\right)=A(s_1 \mid s_2, s_3)=\frac{1}{9} \left(3 \mathcal{G}_0\left(s_1,s_2\right)+3 s_1 \mathcal{G}_1\left(s_1,s_2\right)+\left(-s_1^2+2 s_2 s_1+2 s_2^2\right) \mathcal{G}_2\left(s_1,s_2\right)\right)\,.
\ee
Therefore we can write $\mathcal{C}_{p,q}$ as a combinations of $\widehat{\mathcal{W}}^{(k)}_{p,q}$. For example
\be\label{eq:CpqtoWk}
\begin{array}{l}
 \mathcal{C}_{0,0}=\frac{1}{3} \widehat{\mathcal{W}}_{0,0}^{\text{(0)}}\,,~ \mathcal{C}_{0,1}=\frac{1}{3} \left(\widehat{\mathcal{W}}_{0,0}^{\text{(0)}}-\widehat{\mathcal{W}}_{0,0}^{\text{(1)}}\right) \,,~\mathcal{C}_{0,2}=\frac{1}{27} \left(\widehat{\mathcal{W}}_{0,0}^{\text{(0)}}+9 \widehat{\mathcal{W}}_{1,0}^{\text{(0)}}-3 \widehat{\mathcal{W}}_{0,0}^{\text{(2)}}\right)\,, \mathcal{C}_{0,3}=\frac{1}{3} \left(\widehat{\mathcal{W}}_{1,0}^{\text{(0)}}-\widehat{\mathcal{W}}_{1,0}^{\text{(1)}}\right)\\
 \mathcal{C}_{0,4}=\frac{1}{27} \left(\widehat{\mathcal{W}}_{1,0}^{\text{(0)}}+9 \widehat{\mathcal{W}}_{2,0}^{\text{(0)}}-3 \widehat{\mathcal{W}}_{1,0}^{\text{(2)}}\right) \,,~
 \mathcal{C}_{0,5}=\frac{1}{3} \left(\widehat{\mathcal{W}}_{2,0}^{\text{(0)}}-\widehat{\mathcal{W}}_{2,0}^{\text{(1)}}\right) \,,~
 \mathcal{C}_{0,6}=\frac{1}{27} \left(\widehat{\mathcal{W}}_{2,0}^{\text{(0)}}+9 \widehat{\mathcal{W}}_{3,0}^{\text{(0)}}-3 \widehat{\mathcal{W}}_{2,0}^{\text{(2)}}\right) \\
 \mathcal{C}_{0,7}=\frac{1}{3} \left(\widehat{\mathcal{W}}_{3,0}^{\text{(0)}}-\widehat{\mathcal{W}}_{3,0}^{\text{(1)}}\right) \,,~
 \mathcal{C}_{1,0}=\frac{1}{27} \left(-2 \widehat{\mathcal{W}}_{0,0}^{\text{(0)}}+9 \widehat{\mathcal{W}}_{1,0}^{\text{(0)}}+6 \widehat{\mathcal{W}}_{0,0}^{\text{(2)}}\right) \,,~
 \mathcal{C}_{1,1}=\frac{1}{3} \left(-\widehat{\mathcal{W}}_{0,1}^{\text{(0)}}+\widehat{\mathcal{W}}_{1,0}^{\text{(0)}}-\widehat{\mathcal{W}}_{1,0}^{\text{(1)}}\right) \\
 \mathcal{C}_{1,2}=\frac{1}{9} \left(-3 \widehat{\mathcal{W}}_{0,1}^{\text{(0)}}-\frac{1}{3} \widehat{\mathcal{W}}_{1,0}^{\text{(0)}}+6 \widehat{\mathcal{W}}_{2,0}^{\text{(0)}}+3 \widehat{\mathcal{W}}_{0,1}^{\text{(1)}}+\widehat{\mathcal{W}}_{1,0}^{\text{(2)}}\right) \,,~
 \mathcal{C}_{1,3}=\frac{1}{9} \left(-\frac{1}{3} \widehat{\mathcal{W}}_{0,1}^{\text{(0)}}-3 \widehat{\mathcal{W}}_{1,1}^{\text{(0)}}+6 \widehat{\mathcal{W}}_{2,0}^{\text{(0)}}-6 \widehat{\mathcal{W}}_{2,0}^{\text{(1)}}+\widehat{\mathcal{W}}_{0,1}^{\text{(2)}}\right) \\
 \mathcal{C}_{1,4}=\frac{1}{3} \left(\widehat{\mathcal{W}}_{1,1}^{\text{(1)}}-\widehat{\mathcal{W}}_{1,1}^{\text{(0)}}\right)+\widehat{\mathcal{W}}_{3,0}^{\text{(0)}} \,,~
 \mathcal{C}_{1,5}=-\frac{1}{27} \widehat{\mathcal{W}}_{1,1}^{\text{(0)}}-\frac{1}{3} \widehat{\mathcal{W}}_{2,1}^{\text{(0)}}+\widehat{\mathcal{W}}_{3,0}^{\text{(0)}}-\widehat{\mathcal{W}}_{3,0}^{\text{(1)}}+\frac{1}{9} \widehat{\mathcal{W}}_{1,1}^{\text{(2)}} \\
 \mathcal{C}_{2,0}=\frac{1}{27} \left(-2 \widehat{\mathcal{W}}_{1,0}^{\text{(0)}}+9 \widehat{\mathcal{W}}_{2,0}^{\text{(0)}}+6 \widehat{\mathcal{W}}_{1,0}^{\text{(2)}}\right) \,~
 \mathcal{C}_{2,1}=\frac{1}{27} \left(-9 \left(\widehat{\mathcal{W}}_{1,1}^{\text{(0)}}-\widehat{\mathcal{W}}_{2,0}^{\text{(0)}}+\widehat{\mathcal{W}}_{2,0}^{\text{(1)}}\right)+2 \widehat{\mathcal{W}}_{0,1}^{\text{(0)}}-6 \widehat{\mathcal{W}}_{0,1}^{\text{(2)}}\right) \\
 \mathcal{C}_{2,2}=\frac{1}{3} \left(\widehat{\mathcal{W}}_{0,2}^{\text{(0)}}-\widehat{\mathcal{W}}_{1,1}^{\text{(0)}}-\frac{1}{3} \widehat{\mathcal{W}}_{2,0}^{\text{(0)}}+3 \widehat{\mathcal{W}}_{3,0}^{\text{(0)}}+\widehat{\mathcal{W}}_{1,1}^{\text{(1)}}+\widehat{\mathcal{W}}_{2,0}^{\text{(2)}}\right) \\
 \mathcal{C}_{2,3}=\frac{1}{27} \left(-9 \left(2 \widehat{\mathcal{W}}_{2,1}^{\text{(0)}}-3 \widehat{\mathcal{W}}_{3,0}^{\text{(0)}}+3 \widehat{\mathcal{W}}_{3,0}^{\text{(1)}}\right)+9 \widehat{\mathcal{W}}_{0,2}^{\text{(0)}}+\widehat{\mathcal{W}}_{1,1}^{\text{(0)}}-9 \widehat{\mathcal{W}}_{0,2}^{\text{(1)}}-3 \widehat{\mathcal{W}}_{1,1}^{\text{(2)}}\right) \\
 \mathcal{C}_{3,0}=\frac{1}{27} \left(-2 \widehat{\mathcal{W}}_{2,0}^{\text{(0)}}+9 \widehat{\mathcal{W}}_{3,0}^{\text{(0)}}+6 \widehat{\mathcal{W}}_{2,0}^{\text{(2)}}\right) \,,~
 \mathcal{C}_{3,1}=\frac{1}{27} \left(-9 \left(\widehat{\mathcal{W}}_{2,1}^{\text{(0)}}-\widehat{\mathcal{W}}_{3,0}^{\text{(0)}}+\widehat{\mathcal{W}}_{3,0}^{\text{(1)}}\right)+2 \widehat{\mathcal{W}}_{1,1}^{\text{(0)}}-6 \widehat{\mathcal{W}}_{1,1}^{\text{(2)}}\right) \,.\\
\end{array}
\ee
Using the bounds on $\wmW_{p,q}^{(k)}$, we can get bounds on $\mathcal{C}_{p,q}$. For example first few are shown in the table \eqref{tab:Cpqbounds}
\begin{table}[hbt!]
\centering
\begin{tabular}{|L|L|L||L|L|L||L|L|L|}
\hline
 \text{} & \text{Lower} & \text{Upper} & \text{} & \text{Lower} & \text{Upper} & \text{} & \text{Lower} & \text{Upper} \\
\hline
 \mathcal{C}_{0,4} & -0.0740741 & -0.0271991 & \mathcal{C}_{1,2} & -0.680673 & 0.730324 & \mathcal{C}_{2,1} & -0.373698 & 0.417643 \\
\hline
 \mathcal{C}_{0,5} & -0.046875 & 0.046875 & \mathcal{C}_{1,3} & -0.292318 & 0.283529 & \mathcal{C}_{2,2} & -0.150892 & 0.196345 \\
\hline
 \mathcal{C}_{0,6} & -0.015625 & 0.0118068 & \mathcal{C}_{1,4} & -0.105584 & 0.136287 & \mathcal{C}_{2,3} & -0.166635 & 0.130623 \\
\hline
 \mathcal{C}_{0,7} & -0.0065918 & 0.0065918 & \mathcal{C}_{1,5} & -0.0435567 & 0.0574986 & \mathcal{C}_{3,0} & -0.01043 & 0.0378418 \\
\hline
 \mathcal{C}_{1,1} & -0.375 & 0.1875 & \mathcal{C}_{2,0} & 0.148148 & 0.195023 & \mathcal{C}_{3,1} & -0.0682325 & 0.0558107 \\
\hline
\end{tabular}
\caption{Bounds on $\mathcal{C}_{p,q}$. Note that $\widehat{\mathcal{W}}^{(k)}_{1,0}=1$ in our normalization. Also note that $\mathcal{C}_{p,q}$ which contains the subtraction constants $\widehat{\mathcal{W}}_{0,0}^{(k)}$ can't be bounded. In order to restore the normalization and get the two sided bounds in terms of $\widehat{\mathcal{W}}_{1,0}^{(k)}$ one would have to replace $\widehat{\mathcal{W}}_{p,q}^{(k)}\to\frac{\widehat{\mathcal{W}}_{p,q}^{(k)}}{\widehat{\mathcal{W}}_{1,0}^{(k)}}$ in table \eqref{tab:boundsWk} and use in equation \eqref{eq:CpqtoWk}.}
\label{tab:Cpqbounds}
\end{table}
We have used the normalization $\widehat{\mathcal{W}}^{(k)}_{1,0}=1$. We restore the normalization and get the two sided bounds in terms of $\widehat{\mathcal{W}}_{1,0}^{(k)}$, we replace $\widehat{\mathcal{W}}_{p,q}^{(k)}\to\frac{\widehat{\mathcal{W}}_{p,q}^{(k)}}{\widehat{\mathcal{W}}_{1,0}^{(k)}}$ in table \eqref{tab:boundsWk} and use that in equation \eqref{eq:CpqtoWk}. For example 
\be
\begin{split}
\frac{1}{3} \left(-\frac{1}{8} \widehat{\mathcal{W}}_{1,0}^{\text{(0)}}-\widehat{\mathcal{W}}_{1,0}^{\text{(1)}}\right)&<\mathcal{C}_{1,1}<\frac{1}{3} \left(\frac{25}{16} \widehat{\mathcal{W}}_{1,0}^{\text{(0)}}-\widehat{\mathcal{W}}_{1,0}^{\text{(1)}}\right)\,,\\
\frac{1}{9} \left(\frac{1}{24} (-89) \widehat{\mathcal{W}}_{1,0}^{\text{(0)}}-\frac{270}{79} \widehat{\mathcal{W}}_{1,0}^{\text{(1)}}+\widehat{\mathcal{W}}_{1,0}^{\text{(2)}}\right)&<\mathcal{C}_{1,2}< \frac{1}{9} \left(\frac{211}{96} \widehat{\mathcal{W}}_{1,0}^{\text{(0)}}+\frac{27}{8} \widehat{\mathcal{W}}_{1,0}^{\text{(1)}}+\widehat{\mathcal{W}}_{1,0}^{\text{(2)}}\right)\,.
\end{split}
\ee
If we put $\widehat{\mathcal{W}}^{(k)}_{1,0}=1$, then we recover the results that are given in \eqref{tab:Cpqbounds}.  Using the constraints in the table \eqref{tab:Cpqbounds}, we find the finite region of theory space as depicted in figure \eqref{fig:finiteregionCs}. Some of the known theories are also indicated.

For the 2-loop chiral perturbation theory and the $O(3)$ Lovelace-Shapiro model, we have calculated some of the $\wmW^{(k)}_{p,q}$ and $\mathcal{C}_{p,q}$ with our normalization $\wmW^{(k)}_{1,0}=1$, which are presented in the appendix \eqref{ap:knowntheoriesbounds}.

\section{Summary and future directions}

In this paper, we have applied the techniques in geometric function theory, namely typically real functions to 2-2 scattering amplitudes with global $O(N)$ symmetry. We have used the fully crossing symmetric dispersion relation as in \cite{AK, ASAZ}. 
\\
The main results of the paper are the following: 
\begin{itemize}

\item In the case of theory with $O(N)$ global symmetry, there exist three independent sets of fully crossing symmetric combinations ( see \eqref{def:G0}, \eqref{def:G1}, \eqref{def:G2}) of isospin amplitudes, which was studied in \cite{roy}. We have written down \textit{three collections of fully crossing symmetric dispersion relations in $z$-variable} for fixed $a$.  

\item The dispersion relations empowered us to derive \textit{three sets of independent sum rules}. These \textit{new} sum rules arise because of the cancellation of unphysical powers of $x$.

\item We have written down the partial wave expansion for the dispersion relations through partial wave expansion of the isospin amplitudes (see \eqref{eq:DiscG_k_aI}). These partial wave expansions allow us to derive of \textit{three novel sets of positivity conditions} on the Taylor coefficients of amplitudes around the crossing symmetric point (see \eqref{eq:positivityW0}, \eqref{eq:positivityW1}, \eqref{eq:positivityW2}) employing the unitarity and positivity of Gegenbauer polynomials.

\item We find \textit{three kinds of typically real functions for the $O(N)$ model} \eqref{eq:FktoGk}. Practising the dispersion relations and positivity of the absorptive part of the isospin amplitude for a fixed range of $a$, we have shown that these three functions \eqref{eq:FktoGk} exhibit the Robertson representation \eqref{eq:RobRepfinal}. Robertson representation is a necessary and sufficient condition for typically real functions. Hence we have concluded that these three functions \eqref{eq:FktoGk} are typically real. 

\item Typically real-ness of the functions \eqref{eq:FktoGk} enable us to formulate the  Bieberbach-Rogosinski inequalities for typically real functions \eqref{eq:Rogbounds}. 

\item The Bieberbach-Rogosinski inequalities and the positivity conditions permit us to write down two-sided bounds on the three groups of Taylor coefficients for pion scattering (table \eqref{tab:boundsWk}). These \textit{three assortments of bounds allow us to establish bounds on Taylor coefficients of any physical amplitudes around the crossing symmetric point} (see for example table  \eqref{tab:Cpqbounds}). \textit{Our bounds presented in table \eqref{tab:boundsWk}, do not depend on $N$.}

\item We have supported our bounds by comparing them against know theories, namely 2-loop chiral perturbative amplitude with experimental input on the $b_i$ parameters and the $O(3)$ Lovelace-Shapiro model.
\end{itemize}
Here are our immediate future directions:

\begin{itemize}

\item These three sets of dispersion relations can be employed in the matter of CFT Mellin amplitude to make connections with Polyakov Mellin Bootstrap \cite{Pol, ks, usprl, ON, PFKGASAZ}. The upcoming work of \cite{KGAZ}  will discuss the Witten block expansion.  The upcoming work of \cite{Apratim} will show that a crossing antisymmetric correlator could be expanded in a manifestly crossing antisymmetric basis by introducing a crossing antisymmetric dispersion relation. 

\item An exciting application will be to see the implications of our sum rules \eqref{sumrulev1} to the CFT correlators of charged fields discussed in \cite{Ghosh:2021ruh} and the functionals therein. 

\item Another charming area worth exploring is CFT Mellin amplitudes in light of geometric function theory. It will be interesting to relate CFT typically real-ness with swampland conditions considered recently in \cite{Kundu:2021qpi,Caron-Huot:2021enk} as well as the tricky correlator bounds found in \cite{Paulos:2020zxx}. It will be very interesting to relate the geometric function theory analysis to the positive geometry of CFT, discussed in \cite{positivegeo}.

\item It will be appealing to see the implications of our bounds and sum rules \eqref{sumrulev1} to S-matrix bootstrap for pion amplitudes \cite{ABPHAS,ABAS, andrea} and to the dual the S-matrix bootstrap \cite{dual}.

\end{itemize}

\section*{Acknowledgments} 

We thank Aninda Sinha for suggesting the problem and numerous helpful discussions. We thank Kausik Ghosh, Parthiv Haldar, Apratim Kaviraj, Prashanth Raman for their comments on the draft. We also thank Shaswat Tiwari for helping us with the S-matrix bootstrap data.

\appendix

\section{Verification of various formulas against $O(3)$ Lovelace-Shapiro model}\label{ap:LSsumrules}

In this section, we will verify our various formulas for the Lovelace-Shapiro model \cite{ABAS,LS}.
\be\label{LS_model_full}
A(s_1 \mid s_2, s_3)=\frac{\Gamma(1-\beta(s)) \Gamma(1-\beta(t))}{\Gamma(1-\beta(s)-\beta(t))}\,.
\ee
with $\beta(s)=\beta_{0}+\beta^{\prime} s, \quad \beta\left(m_{\rho}^{2}\right)=1\,.$

For demonstration purpose, we will choose $\b_0=\b'=\frac{1}{2}$. 
Lovelace-Shapiro model, in our convention, we write
\be
A^{(LS)}(s_1,s_2)=A^{(LS)}(s_1 \mid s_2, s_3)=\frac{\Gamma \left(\frac{1}{2}-\frac{s_1}{2}\right) \Gamma \left(\frac{1}{2}-\frac{s_2}{2}\right)}{\Gamma \left(-\frac{s_1}{2}-\frac{s_2}{2}\right)}\,,
\ee
and
\be\label{LS_isoM}
\begin{split}
&\mathcal{M}^{(0)}_{(LS)}(s_1, s_2)=\frac{3}{2}\left(A^{(LS)}(s_1, s_2)+A^{(LS)}(s_1, s_3)\right)-\frac{1}{2} A^{(LS)}(s_2, s_3), \\
& \mathcal{M}^{(1)}_{(LS)}(s_1,s_2)=A^{(LS)}(s_1, s_2)-A^{(LS)}(s_1, s_3), \quad \mathcal{M}^{(2)}_{(LS)}(s_1, s_2)=A^{(LS)}(s_2, s_3)\,.
\end{split}
\ee

\subsection{Dispersion relation}
In this section, we verify the dispersion integral \eqref{eq:disper00} against Lovelace-Shapiro model. We plug \eqref{LS_isoM} in \eqref{def:G0}, \eqref{def:G1}, \eqref{def:G2}, to get $\mathcal{G}_k(s_1,s_2)$. We calculate the discontinuity $\text{Disc}\mathcal{G}_k(s_1,s_2)$. Then we plug back that calculated $\text{Disc}\mathcal{G}_k(s_1,s_2)$ in formula \eqref{eq:disper00} and check if the LHS is equal to RHS \textit{i.e.} if the dispersion relations are correct. The subtraction constants $\a_0^{(k)}$ are given by
\be
\a_0^{(0)}=0\,,\a_0^{(1)}=-\frac{3 \pi}{2} \,,\a_0^{(2)}=\frac{-3 \pi  \log (4)}{4}\,.
\ee
The $s$-channel discontinuities are given by
\be\label{eq:discG0LS}
\text{Disc}\mathcal{G}_0(s_1,s_2^{(+)}(s_1,a))=\sum_{k=0}^{\infty}\pi \d(s_1-2k-1)\frac{(-1)^k \left(-\frac{\Gamma \left(\frac{1}{4} (-2 k (\lambda -1)-\lambda +3)\right)}{\Gamma \left(-\frac{1}{4} (2 k+1) (\lambda +1)\right)}-\frac{\Gamma \left(\frac{1}{4} (\lambda +2 k (\lambda +1)+3)\right)}{\Gamma \left(\frac{1}{4} (2 k+1) (\lambda -1)\right)}\right)}{k!}\,,
\ee

\be\label{eq:discG1LS}
\begin{split}
&\text{Disc}\mathcal{G}_1(s_1,s_2^{(+)}(s_1,a))=\sum_{k=0}^{\infty}\pi \d(s_1-2k-1)\frac{3 (-1)^k \lambda  (a-2 k-1) }{(2 k+1) (3 a+2 k+1) (-3 a+4 k+2) \Gamma (k+1)}\\
&\left(\frac{((2 k+1) (\lambda +1)-a (\lambda +3)) \Gamma \left(\frac{1}{4} (-2 k (\lambda -1)-\lambda +3)\right)}{\Gamma \left(-\frac{1}{4} (2 k+1) (\lambda +1)\right)}+\frac{((2 k+1) (\lambda -1)-a (\lambda -3)) \Gamma \left(\frac{1}{4} (\lambda +2 k (\lambda +1)+3)\right)}{\Gamma \left(\frac{1}{4} (2 k+1) (\lambda -1)\right)}\right)\,,
\end{split}
\ee
\be\label{eq:discG2LS}
\begin{split}
&\text{Disc}\mathcal{G}_2(s_1,s_2^{(+)}(s_1,a))=\sum_{k=0}^{\infty}\pi \d(s_1-2k-1)\frac{12 (-1)^k \left(-\frac{\Gamma \left(\frac{1}{4} (-2 k (\lambda -1)-\lambda +3)\right)}{(\lambda +3) \Gamma \left(-\frac{1}{4} (2 k+1) (\lambda +1)\right)}-\frac{\Gamma \left(\frac{1}{4} (\lambda +2 k (\lambda +1)+3)\right)}{(\lambda -3) \Gamma \left(\frac{1}{4} (2 k+1) (\lambda -1)\right)}\right)}{(2 k+1)^2 k! \sqrt{\frac{4 a}{-a+2 k+1}+1}}\,,
\end{split}
\ee
with 
\be
\l=\sqrt{\frac{4 a}{-a+2 k+1}+1}\,.
\ee
We put $\text{Disc}\mathcal{G}_k(s_1,s_2)$ given in the above formulae in the dispersion relation \eqref{eq:disper00}. For numerical illustration, we truncate the $k$-sum at $k=k_{\max}$. We note that, because of $\d(s_1-2k-1)$, the $s_1'$ integral can be easily done. The comparison is given the table below \eqref{tab:LSmodel_dispersion}

\begin{table}[hbt!]
\centering
\begin{tabular}{|l|l|l|l|}
\hline
 $\mathcal{G}_k\left(s_1,s_2\right)$ & \text{Exact} & $k_{\max }\text{=100}$ & $k_{\max }\text{=150}$  \\
\hline
$ \mathcal{G}_0\left(s_1=\frac{1}{22},s_2=\frac{i}{13}\right)$ & 0.0038224\, +0.0035656 i & 0.00382198\, +0.0035653 i & 0.0038222\, +0.0035654 i \\
\hline
 $\mathcal{G}_0\left(s_1=\frac{1}{2}-i,s_2=1+\frac{i}{3}\right)$ & 0.939177\, -0.790044 i & 0.935403\, -0.796079 i & 0.936207\, -0.794022 i  \\
\hline
$ \mathcal{G}_1\left(s_1=\frac{1}{22},s_2=\frac{i}{13}\right)$ & -4.70217+0.0097187 i & -4.70217+0.0097187 i & -4.70217+0.0097187 i  \\
\hline
$ \mathcal{G}_1\left(s_1=\frac{1}{2}-i,s_2=1+\frac{i}{3}\right) $& 1.20737\, -4.67827 i & 1.20735\, -4.67832 i & 1.20736\, -4.67829 i  \\
\hline
 $\mathcal{G}_2\left(s_1=\frac{1}{22},s_2=\frac{i}{13}\right)$ & -3.25556+0.0100111 i & -3.25556+0.0100111 i & -3.25556+0.0100111 i \\
\hline
$ \mathcal{G}_2\left(s_1=\frac{1}{2}-i,s_2=1+\frac{i}{3}\right)$ & -2.60411-2.07626 i & -2.60411-2.07626 i & -2.60411-2.07626 i  \\
\hline
\end{tabular}
\caption{Verification of the dispersion integral \eqref{eq:disper00} against Lovelace-Shapiro model. We put $\text{Disc}\mathcal{G}_k(s_1,s_2)$ from \eqref{eq:discG0LS}, \eqref{eq:discG1LS}, \eqref{eq:discG2LS} in formula \eqref{eq:disper00}. For numerical purpose, we truncate the $k$-sum upto say $k_{\max}$. }
\label{tab:LSmodel_dispersion}
\end{table}

\subsection{Inversion formulas and Sum rules}
From the formulas \eqref{LS_isoM}, we can calculate the $s$-channel discontinuity $\mathcal{A}^{(I)}_{(LS)}(s_1,s_2)=Disc \left[\mathcal{M}^{(I)}_{(LS)}(s_1,s_2)\right]$.
\be
\begin{split}
&\mathcal{A}^{(0)}_{(LS)}(s_1,s_2)=\sum_{k=0}^{\infty}\pi\d(s_1-2k-1)\frac{3 (-1)^k \left(-\frac{\Gamma \left(\frac{1}{2}-\frac{1}{4} (2 k+1) (\cos (\theta )-1)\right)}{\Gamma \left(-\frac{1}{4} (2 k+1) (\cos (\theta )+1)\right)}-\frac{\Gamma \left(\frac{1}{4} (2 \cos (\theta ) k+2 k+\cos (\theta )+3)\right)}{\Gamma \left(\frac{1}{4} (2 k+1) (\cos (\theta )-1)\right)}\right)}{k!}\,,\\
&\mathcal{A}^{(1)}_{(LS)}(s_1,s_2)=\sum_{k=0}^{\infty}\pi\d(s_1-2k-1)\frac{2 (-1)^k \left(\frac{\Gamma \left(\frac{1}{4} (2 \cos (\theta ) k+2 k+\cos (\theta )+3)\right)}{\Gamma \left(\frac{1}{4} (2 k+1) (\cos (\theta )-1)\right)}-\frac{\Gamma \left(\frac{1}{2}-\frac{1}{4} (2 k+1) (\cos (\theta )-1)\right)}{\Gamma \left(-\frac{1}{4} (2 k+1) (\cos (\theta )+1)\right)}\right)}{k!}\,,\\
&\mathcal{A}^{(2)}_{(LS)}(s_1,s_2)=0\,,
\end{split}
\ee
where $\cos(\theta)=1+\frac{2s_2}{s_1}$. We will work in $d=4$ for illustration purpose. Now one can calculate the partial wave coefficients
\be\label{aell_LS}
\Phi(s_1;\a=1/2)a_\ell^{(I)}(s_1)=\frac{1}{2} \int_{-1}^1 P_{\ell}(\cos (\theta )) \mathcal{A}^{(I)}_{(LS)}\left(s_1,s_2=s_1\frac{\cos(\theta)-1}{2}\right) \, d\cos (\theta )\,.
\ee
We will truncate $\ell$-sum to $L_{\max}$ and $k$-sum to $k_{\max}$. Knowing the $a_\ell^{(I)}(s_1)$, we can calculate the $\mathcal{W}^{(k)}_{p,q}$ via formula \eqref{eq:invWk}. Due to the $\d(s_1-2k-1)$, the $s_1$ integrals in \eqref{eq:invWk} can be easily done. Putting the equations \eqref{LS_isoM} in \eqref{def:G0}, \eqref{def:G1}, \eqref{def:G2} to get $\mathcal{G}_k(s_1,s_2)$, we can calculate the $\mathcal{W}^{(k)}_{p,q}$ directly after expanding the $\mathcal{G}_k(s_1,s_2)$ in powers of $x,y$. We can compare the directly calculated $\mathcal{W}^{(k)}_{p,q}$ with the inversion formula \eqref{eq:invWk}. The comparison is shown in the table \eqref{tab:LSinv} below
\begin{table}[hbt!]
\centering
\begin{tabular}{|c |c|c|c|c|}
\hline
 $\mathcal{W}_{2,1}^{\text{(k)}}$ & Exact & $L_{\max }\text{=6}$ & $L_{\max }\text{=10}$ & $L_{\max }\text{=20}$ \\
\hline
 $\mathcal{W}_{2,1}^{\text{(0)}}$ & -2.49215 & -2.49217 & -2.49215 & -2.49215 \\
\hline
 $\mathcal{W}_{2,1}^{\text{(1)}}$ & -8.99964 & -8.99965 & -8.99964 & -8.99964 \\
\hline
 $\mathcal{W}_{2,1}^{\text{(2)}}$ & -6.00331 & -6.00331 & -6.00331 & -6.00331 \\
\hline
\end{tabular}
\caption{$\mathcal{W}^{(k)}_{2,1}$ calculated directly using \eqref{LS_isoM} in \eqref{def:G0}, \eqref{def:G1}, \eqref{def:G2}, compare with calculated using formula \eqref{eq:invWk} and \eqref{aell_LS}. We have truncated the $k$-sum to $k_{\max}=40$ and $\ell$-sum to $L_{\max}$. }
\label{tab:LSinv}
\end{table}

\begin{figure}[hbt!]
\centering
\includegraphics[scale=0.7]{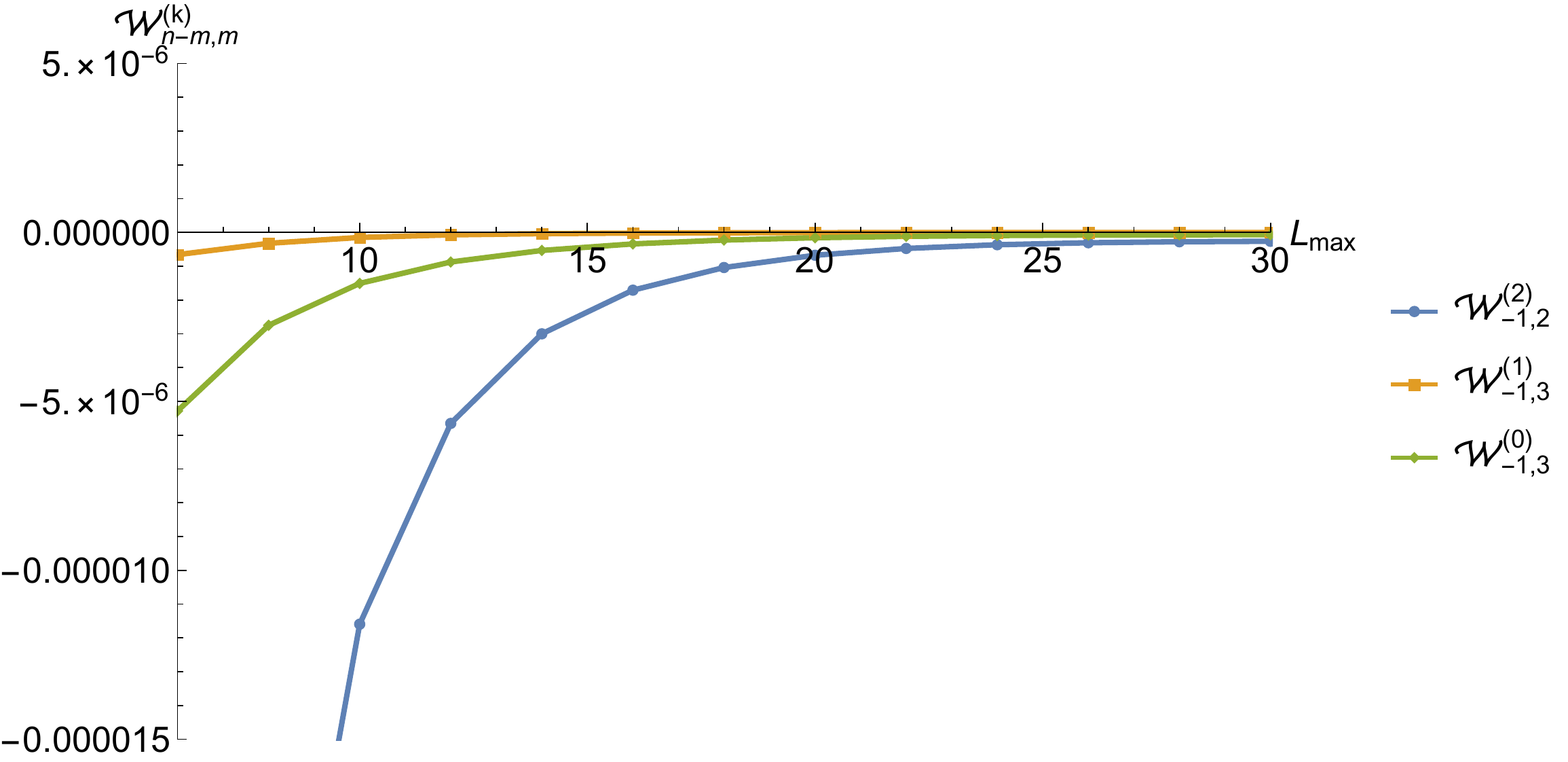}
\caption{We calculate $\mathcal{W}^{(k)}_{n-m,m}$ for $m>n$, using formula \eqref{eq:invWk} and \eqref{aell_LS}. The $\mathcal{W}^{(k)}_{n-m,m}$ for $m>n$ should vanish.  We have truncated the $k$-sum to $k_{\max}=40$ and $\ell$-sum to $L_{\max}$. We see that as we increase $L_{\max}$ the $\mathcal{W}_{-1,3}^{(0)}, \mathcal{W}_{-1,3}^{(0)}, \mathcal{W}_{-1,2}^{(2)}$ go toward zero.}
\label{fig:LSsumrule}
\end{figure}

We can calculate $\mathcal{W}^{(k)}_{n-m,m}$ for $m>n$, the procedure is exactly same. Now since for $m>n$ which implies $(y/x)^m x^n$ \textit{i.e} negative powers of $x$. There should not be any negative powers of $x$ as in the expression \eqref{eq:GktoWk} due to locality. The coefficients $\mathcal{W}^{(k)}_{n-m,m}$ for $m>n$ should vanish. For  Lovelace-Shapiro model, we have shown this in the figure \eqref{fig:LSsumrule} below

\section{Positivity conditions on of $\mathcal{W}_{n,m}^{(1)},\mathcal{W}_{n,m}^{(2)}$}\label{ap:positivity}
\subsection{Finding $\Upsilon _{n,r,m}^{(1)}(\mu, \delta),\Upsilon _{n,r,m}^{(2)}(\mu, \delta),\chi _{n,r,m}^{(1)}(\mu, \delta),\chi _{n,r,m}^{(2)}(\mu, \delta)$}
First, we take the combinations
\be
T_1=\sum_{r=0}^{m}\left(\Upsilon _{n,r,m}^{(1)}(\mu, \delta)\mathcal{W}_{n-r,r}^{(1)}+\bar{\chi}_{n,r,m}^{(1)}(\mu, \delta)\mathcal{W}_{n-r,r}^{(0)}\right)\,,
\ee
and
\be
T_2=\sum_{r=0}^{m}\left(\Upsilon _{n,r,m}^{(2)}(\mu, \delta)\mathcal{W}_{n-r,r}^{(2)}+\bar{\chi}_{n,r,m}^{(2)}(\mu, \delta)\mathcal{W}_{n-r,r}^{(0)}\right)\,.
\ee
From $T_1, T_2$ we collect all the coefficients of $a_{\ell}^{(I)}$ after we put the formula for $\mathcal{W}_{n-m,m}^{(k)}$ from the inversion formula \eqref{eq:invWk}. We will change the variable $s_1=\d+\frac{2\m}{3}$. We always choose $\Upsilon _{n,m,m}^{(k)}(\mu, \delta)=1, ~~k=1,2\,.$

For $T_1$, we collect the coefficients of $a_{\ell}^{(0)}C_{\ell-i}^{(\alpha +i)}\left(\frac{2 \mu }{3 \delta }+1\right)$, $a_{\ell}^{(1)}C_{\ell-i}^{(\alpha +i)}\left(\frac{2 \mu }{3 \delta }+1\right)$ for $i=0,1,2,3,\dots m-1$ and $a_{\ell}^{(2)}C_{\ell-m}^{(\alpha +m)}\left(\frac{2 \mu }{3 \delta }+1\right)$. Putting these coefficients to zero give solutions to $\Upsilon _{n,r,m}^{(1)}(\mu, \delta),\bar{\chi}_{n,r,m}^{(1)}(\mu, \delta)$. After that we put these solutions in $T_1$. Then we consider $\sum_{r=0}^{m-1}\hat{\chi}_{n,r,m}^{(1)}(\mu, \delta)\mathcal{W}_{n-r,r}^{(0)}+T_1$ and put the coefficients of $a_{\ell}^{(2)}C_{\ell-i}^{(\alpha +i)}\left(\frac{2 \mu }{3 \delta }+1\right)$ for $i=0,1,2,3,\dots m-1$ to zero to get solutions for $\hat{\chi}_{n,r,m}^{(1)}(\mu, \delta)$. Now full solutions, we already have $\Upsilon _{n,r,m}^{(1)}(\mu, \delta)$, and ${\chi}_{n,r,m}^{(1)}(\mu, \delta)=\bar{\chi}_{n,r,m}^{(1)}(\mu, \delta)+\hat{\chi}_{n,r,m}^{(1)}(\mu, \delta)$. For example, we have worked out $m=1$ case
\be
\Upsilon _{n,0,1} \mathcal{W}_{n,0}^{\text{(1)}}+\mathcal{W}_{n,1}^{\text{(1)}}+\bar{\chi }_{n,0,1} \mathcal{W}_{n,0}^{\text{(0)}}+\bar{\chi }_{n,1,1} \mathcal{W}_{n,1}^{\text{(0)}}\,,
\ee
\begin{scriptsize}
\be
\begin{split}
T_1=&\frac{2  \bar{\chi }_{n,0,1} C_{\ell }^{(\alpha )}\left(\frac{2 \mu }{3 \delta }+1\right) \left(a_{\ell }^{\text{(0)}}\left(s_1\right)+(N-1) a_{\ell }^{\text{(2)}}\left(s_1\right)\right)}{\pi  N \left(\delta +\frac{2 \mu }{3}\right)^{2 n+1}}+\frac{3 \Upsilon _{n,0,1} C_{\ell }^{(\alpha )}\left(\frac{2 \mu }{3 \delta }+1\right) \left(2 a_{\ell }^{\text{(0)}}\left(s_1\right)+N a_{\ell }^{\text{(1)}}\left(s_1\right)-(N+2) a_{\ell }^{\text{(2)}}\left(s_1\right)\right)}{2 \pi  N \left(\delta +\frac{2 \mu }{3}\right)^{2 (n+1)}}\\
&+\frac{ \bar{\chi }_{n,1,1} \left(a_{\ell }^{\text{(0)}}\left(s_1\right)+(N-1) a_{\ell }^{\text{(2)}}\left(s_1\right)\right) \left(8 \alpha  (3 \delta +2 \mu ) C_{\ell -1}^{(\alpha +1)}\left(\frac{2 \mu }{3 \delta }+1\right)-3 \delta  (2 n+1) C_{\ell }^{(\alpha )}\left(\frac{2 \mu }{3 \delta }+1\right)\right)}{\pi  \delta  N 3 \left(\delta +\frac{2 \mu }{3}\right)^{2 (n+1)}}\\
&+\frac{3  \left(\frac{4 \alpha  \left(\delta +\frac{2 \mu }{3}\right) C_{\ell -1}^{(\alpha +1)}\left(\frac{2 \mu }{3 \delta }+1\right) \left(2 a_{\ell }^{\text{(0)}}\left(s_1\right)+N a_{\ell }^{\text{(1)}}\left(s_1\right)-(N+2) a_{\ell }^{\text{(2)}}\left(s_1\right)\right)}{\delta }-C_{\ell }^{(\alpha )}\left(\frac{2 \mu }{3 \delta }+1\right) \left(2 n a_{\ell }^{\text{(0)}}\left(s_1\right)+(n+4) N a_{\ell }^{\text{(1)}}\left(s_1\right)-n (N+2) a_{\ell }^{\text{(2)}}\left(s_1\right)\right)\right)}{2 \pi  N \left(\delta +\frac{2 \mu }{3}\right)^{2 n+3}}\,.
\end{split}
\ee
\end{scriptsize}
Following the method described above, we get 
\be
\Upsilon _{n,0,1}=\frac{3 (n+4)}{3 \delta +2 \mu },~\bar{\chi }_{n,0,1}=\frac{27 (2 n (N+2)-15 N+18)}{8 (N-1) (3 \delta +2 \mu )^2},~\bar{\chi }_{n,1,1}=\frac{9 (N+2)}{4 (N-1) (3 \delta +2 \mu )}\,.
\ee
Now we add $\hat{\chi}_{n,0,1}^{(1)}(\mu, \delta)\mathcal{W}_{n,0}^{(0)}+T_1$ and put the coefficients of $a_{\ell}^{(2)}C_{\ell}^{(\alpha )}\left(\frac{2 \mu }{3 \delta }+1\right)$ to zero. We get
\be
\hat{\chi }_{n,0,1}=\frac{81 N}{(N-1) (3 \delta +2 \mu )^2}\,.
\ee
Then the full solution
\be
\chi _{n,0,1}(\mu, \delta)=\frac{27 (2 n+9) (N+2)}{8 (N-1) (3 \delta +2 \mu )^2},~\Upsilon _{n,0,1}(\mu, \delta)=\frac{3 (n+4)}{3 \delta +2 \mu },~\chi _{n,1,1}(\mu, \delta)=\frac{9 (N+2)}{4 (N-1) (3 \delta +2 \mu )}\,.
\ee
Identical calculations can be repeated for higher values of $m$. 

Similarly for $T_2$, we collect the coefficients of $a_{\ell}^{(2)}C_{\ell-i}^{(\alpha +i)}\left(\frac{2 \mu }{3 \delta }+1\right)$, $a_{\ell}^{(1)}C_{\ell-i}^{(\alpha +i)}\left(\frac{2 \mu }{3 \delta }+1\right)$ for $i=0,1,2,3,\dots m-1$ and $a_{\ell}^{(0)}C_{\ell-m}^{(\alpha +m)}\left(\frac{2 \mu }{3 \delta }+1\right)$. Putting these coefficients to zero give $\Upsilon _{n,r,m}^{(2)}(\mu, \delta),\bar{\chi}_{n,r,m}^{(2)}(\mu, \delta)$. Then we put these solutions in $T_2$. Now we consider $\sum_{r=0}^{m-1}\hat{\chi}_{n,r,m}^{(2)}(\mu, \delta)\mathcal{W}_{n-r,r}^{(0)}+T_2$ and put $a_{\ell}^{(0)}C_{\ell-i}^{(\alpha +i)}\left(\frac{2 \mu }{3 \delta }+1\right)$ for $i=0,1,2,3,\dots m-1$ to zero to get solutions for $\hat{\chi}_{n,r,m}^{(2)}(\mu, \delta)$. Now full solutions, we already have $\Upsilon _{n,r,m}^{(2)}(\mu, \delta)$, and ${\chi}_{n,r,m}^{(2)}(\mu, \delta)=\bar{\chi}_{n,r,m}^{(2)}(\mu, \delta)+\hat{\chi}_{n,r,m}^{(2)}(\mu, \delta)$.

\subsection{Proof of positivity conditions on of $\mathcal{W}_{n,m}^{(1)},\mathcal{W}_{n,m}^{(2)}$}
After finding $\Upsilon _{n,r,m}^{(1)}(\mu, \delta),\Upsilon _{n,r,m}^{(2)}(\mu, \delta),\chi _{n,r,m}^{(1)}(\mu, \delta),\chi _{n,r,m}^{(2)}(\mu, \delta)$, one can check that coefficients of $a_{\ell}^{(I)}C_{\ell-i}^{(\alpha +i)}\left(\frac{2 \mu }{3 \delta }+1\right)$ for $i=0,1,2,3,\dots m$ all are positive in \textit{i.e.} \eqref{eq:positivityW1}, \eqref{eq:positivityW2}

For example, lets consider \eqref{eq:positivityW1} with $m=1$.
\be
\begin{split}
&\frac{ \left(3 \left(\delta -\delta _0\right) \left(3 \delta _0+2 \mu \right) ((2 n+11) N+6)+3 \left(\delta -\delta _0\right){}^2 (2 n+9) (N+2)+8 N \left(3 \delta _0+2 \mu \right){}^2\right)}{4 \pi  (N-1) N \left( \delta _0+\frac{2 \mu }{3}\right)^2 \left(\delta +\frac{2 \mu }{3}\right)^{2 n+3}}a_\ell^{(0)}C_{\ell}^{(\alpha )}\left(\frac{2 \mu }{3 \delta }+1\right)\\
&+\frac{2 \alpha  9^{n+2} (3 \delta +2 \mu )^{-2 (n+1)} \left(N \left(3 \delta _0+2 \mu \right)+\left(\delta -\delta _0\right) (N+2)\right)}{\pi  \delta  (N-1) N \left(3 \delta _0+2 \mu \right)}a_\ell^{(0)}C_{\ell-1}^{(\alpha +1)}\left(\frac{2 \mu }{3 \delta }+1\right)\\
&\frac{\left(\delta -\delta _0\right) 3^{2 n+5} (n+4) (3 \delta +2 \mu )^{-2 n-3}}{2 \pi  \left(3 \delta _0+2 \mu \right)}a_\ell^{(1)}C_{\ell}^{(\alpha )}\left(\frac{2 \mu }{3 \delta }+1\right)+\frac{6 \alpha  \left(\delta +\frac{2 \mu }{3}\right)^{-2 (n+1)}}{\pi  \delta }a_\ell^{(1)}C_{\ell-1}^{(\alpha +1)}\left(\frac{2 \mu }{3 \delta }+1\right)\\
&   \frac{\left(\delta -\delta _0\right) 9^{n+3} (N+2) (3 \delta +2 \mu )^{-2 n-3} \left(9 \delta _0+6 \mu +\left(\delta -\delta _0\right) (2 n+9)\right)}{4 \pi  N \left(3 \delta _0+2 \mu \right){}^2}  a_\ell^{(2)}C_{\ell}^{(\alpha )}\left(\frac{2 \mu }{3 \delta }+1\right)      \\
&  +\frac{2 \alpha  \left(\delta -\delta _0\right) 9^{n+2} (N+2) (3 \delta +2 \mu )^{-2 (n+1)}}{\pi  \delta  N \left(3 \delta _0+2 \mu \right)}a_\ell^{(2)}C_{\ell-1}^{(\alpha +1)}\left(\frac{2 \mu }{3 \delta }+1\right)\geq 0\,.
\end{split}
\ee
The above is always positive for $\d\geq \d_0\geq 0,\a \geq 0,\m\geq 0$. Similarly one can check for any $m$ that equation \eqref{eq:positivityW1} from the fact that coefficients of $a_{\ell}^{(I)}C_{\ell-i}^{(\alpha +i)}\left(\frac{2 \mu }{3 \delta }+1\right)$ for $i=0,1,2,3,\dots m$ are all positive. 

Similar arguments holds for \eqref{eq:positivityW2}. The coefficients of $a_{\ell}^{(0)}C_{\ell-i}^{(\alpha +i)}\left(\frac{2 \mu }{3 \delta }+1\right)$ for $i=0,1,2,3,\dots m$ and coefficients of $a_{\ell}^{(2)}C_{\ell-i}^{(\alpha +i)}\left(\frac{2 \mu }{3 \delta }+1\right)$ for $i=0,1,2,3,\dots m$ are always positive. But for coefficients of $a_{\ell}^{(1)}$, we can check numerically that they are indeed positive (\textit{i.e.} the total sum of the all Gegenbauer polynomials, not individual coefficients of  $C_{\ell-i}^{(\alpha +i)}\left(\frac{2 \mu }{3 \delta }+1\right)$ for $i=0,1,2,3,\dots m$ ).

\section{Positivity of $\text{Disc}\left[ \left(1-\frac{2 \delta _{2,k}}{3}\right)\mathcal{G}_0+\delta _{1,k}\mathcal{G}_1 +\delta _{2,k}\mathcal{G}_2 \right]\left(s_1^{\prime} ; s_2^{(+)}\left(s_1^{\prime} ,a\right)\right)$}\label{ap:positivityofF}

In this section, we prove that $\text{Disc}\left[ \left(1-\frac{2 \delta _{2,k}}{3}\right)\mathcal{G}_0+\delta _{1,k}\mathcal{G}_1 +\delta _{2,k}\mathcal{G}_2 \right]\left(s_1^{\prime} ; s_2^{(+)}\left(s_1^{\prime} ,a\right)\right)>0$ for $s_1'\geq \frac{2\mu}{3},~ N\geq 3,~\mu\geq 4$ and range of $a$ in case of $k=0$ is given by $-\frac{2\m}{9}<a<\frac{2\mu}{3}$, while in case of $k=1,2$ range of $a$ is given by $-\frac{2\m}{9}<a<\frac{2\mu}{9}$. We will be using the positivity of the absorptive part of isospin amplitudes, namely
$
\mathcal{A}^{(I)}\left(s_1^{\prime} ; s_2^{(+)}\left(s_1^{\prime} ,a\right)\right)
$ is non-negative for $-\frac{2\m}{9}<a<\frac{2\mu}{3},~s_1'\geq \frac{2\mu}{3}$. This follows from the partial wave expansion \eqref{eq:IsoA_wave}, \eqref{eq:positiveaell} and the positivity of the Gegenbauer polynomials, 
\be
C_{\ell}^{(\alpha)}\left(\sqrt{\xi\left(s_{1}, a\right)}\right)>0 \text { if } -\frac{2\m}{9}<a<\frac{2\mu}{3},~s_1'\geq \frac{2\mu}{3}\,.
\ee
This is true because $\sqrt{\xi\left(s_{1}, a\right)}>1$ for $-\frac{2\m}{9}<a<\frac{2\mu}{3},~s_1'\geq \frac{2\mu}{3}\,.$

\subsubsection*{$k=0$ case:}
For $k=0$, we have 
\be
\text{Disc}\mathcal{G}_0\left(s_1^{\prime} ; s_2^{(+)}\left(s_1^{\prime} ,a\right)\right)=\frac{\mathcal{A}^{\text{(0)}}\left(s_1^{\prime} ; s_2^{(+)}\left(s_1^{\prime} ,a\right)\right)+(N-1) \mathcal{A}^{\text{(2)}}\left(s_1^{\prime} ; s_2^{(+)}\left(s_1^{\prime} ,a\right)\right)}{N}\,.
\ee
Therefore from non-negativity of $
\mathcal{A}^{(I)}\left(s_1^{\prime} ; s_2^{(+)}\left(s_1^{\prime} ,a\right)\right)
$, we find that $\text{Disc}\mathcal{G}_0\left(s_1^{\prime} ; s_2^{(+)}\left(s_1^{\prime} ,a\right)\right)$ is non-negative for $-\frac{2\m}{9}<a<\frac{2\mu}{3},~s_1'\geq \frac{2\mu}{3}\,.$

\subsubsection*{$k=1$ case:}
For $k=1$, we have 
\be
\begin{split}
&\text{Disc}[\mathcal{G}_0+\mathcal{G}_1]\left(s_1^{\prime} ; s_2^{(+)}\left(s_1^{\prime} ,a\right)\right)=\left[\frac{\frac{1}{2 s_1'-3 a}+\frac{1}{s_1'}+1}{N}\right]\mathcal{A}^{(0)}\left(s_1^{\prime} ; s_2^{(+)}\left(s_1^{\prime} ,a\right)\right)\\
&+\left[\frac{3 \left(3 a-s_1'\right) \sqrt{1-\frac{4 a}{3 a+s_1'}}}{2 s_1' \left(3 a-2 s_1'\right)}\right]\mathcal{A}^{(1)}\left(s_1^{\prime} ; s_2^{(+)}\left(s_1^{\prime} ,a\right)\right)+\left[\frac{\frac{N+2}{3 a-2 s_1'}-\frac{N+2}{s_1'}+2 N-2}{2 N}\right]\mathcal{A}^{(2)}\left(s_1^{\prime} ; s_2^{(+)}\left(s_1^{\prime} ,a\right)\right)\,.
\end{split}
\ee

\subsubsection*{$k=2$ case:}
For $k=2$ we have 
\be
\begin{split}
&\text{Disc}\left[\frac{\mathcal{G}_0}{3}+\mathcal{G}_2\right]\left(s_1^{\prime} ; s_2^{(+)}\left(s_1^{\prime} ,a\right)\right)=\left[-\frac{-3 a \left(s_1'\right){}^2+9 a+2 \left(s_1'\right){}^3-9 s_1'}{9 a N \left(s_1'\right){}^2-6 N \left(s_1'\right){}^3}\right]\mathcal{A}^{(0)}\left(s_1^{\prime} ; s_2^{(+)}\left(s_1^{\prime} ,a\right)\right)\\
&+\left[\frac{9 \left(s_1'-a\right){}^{3/2}}{2 \left(s_1'\right){}^2 \sqrt{3 a+s_1'} \left(2 s_1'-3 a\right)}\right]\mathcal{A}^{(1)}\left(s_1^{\prime} ; s_2^{(+)}\left(s_1^{\prime} ,a\right)\right)+\left[\frac{\frac{3 (N+2)}{2 s_1'-3 a}+\frac{3 (N+2)}{s_1'}+2 (N-1) s_1'}{6 Ns_1'}\right]\mathcal{A}^{(2)}\left(s_1^{\prime} ; s_2^{(+)}\left(s_1^{\prime} ,a\right)\right)\,.
\end{split}
\ee

In both of the cases $k=1,2$, one can see that each coefficients of $\mathcal{A}^{(I)}\left(s_1^{\prime} ; s_2^{(+)}\left(s_1^{\prime} ,a\right)\right)$ are positive if we consider the range $-\frac{2\m}{9}<a<\frac{2\mu}{9},~s_1'\geq \frac{2\mu}{3},~ N\geq 3,~\mu\geq 4$. Once we see all coefficients are positive. Then using the non-negativity of the $\mathcal{A}^{(I)}\left(s_1^{\prime} ; s_2^{(+)}\left(s_1^{\prime} ,a\right)\right)$. We conclude that $k=1,2$ cases are non-negativity for $s_1'\geq \frac{2\mu}{3},~ N\geq 3,~\mu\geq 4$ and range of $a$ is given by $-\frac{2\m}{9}<a<\frac{2\mu}{9}$.

\section{Verifications of bounds with known theories}\label{ap:knowntheoriesbounds}

\subsection{The 2-loop Chiral perturbation theory}
In the 2-loop chiral perturbation theory, we use the experimental values for the parameters in the amplitude (we use the amplitude given in  \cite{Wang:2020jxr, Bijnens:1995yn}).
\be
\begin{split}
&b_1=-0.0785239,~b_2=0.0747244,~b_3=-0.00208975,\\
&b_4=0.0046861,~b_5=0.000143563,~b_6=0.0000942385
\end{split}
\ee
We find the coefficients
\begin{scriptsize}
\be\label{eq:WChiPT}
\begin{split}
&\frac{\widehat{\mathcal{W}}_{0,1}^{\text{(0)}}}{\widehat{\mathcal{W}}_{1,0}^{(0)}}=-0.16534,~\frac{\widehat{\mathcal{W}}_{0,2}^{\text{(0)}}}{\widehat{\mathcal{W}}_{1,0}^{(0)}}=0.00156937,~\frac{\widehat{\mathcal{W}}_{1,1}^{\text{(0)}}}{\widehat{\mathcal{W}}_{1,0}^{(0)}}=-0.0102986,~\frac{\widehat{\mathcal{W}}_{2,0}^{\text{(0)}}}{\widehat{\mathcal{W}}_{1,0}^{(0)}}=0.0182756,~\frac{\widehat{\mathcal{W}}_{3,0}^{\text{(0)}}}{\widehat{\mathcal{W}}_{1,0}^{(0)}}=0.00104625\,,\\
&\frac{\widehat{\mathcal{W}}_{0,1}^{\text{(1)}}}{\widehat{\mathcal{W}}_{1,0}^{(1)}}=-0.164701,~\frac{\widehat{\mathcal{W}}_{0,2}^{\text{(1)}}}{\widehat{\mathcal{W}}_{1,0}^{(1)}}=0.00166561,~\frac{\widehat{\mathcal{W}}_{1,1}^{\text{(1)}}}{\widehat{\mathcal{W}}_{1,0}^{(1)}}=-0.0111136,~\frac{\widehat{\mathcal{W}}_{2,0}^{\text{(1)}}}{\widehat{\mathcal{W}}_{1,0}^{(1)}}=0.0200637,~\frac{\widehat{\mathcal{W}}_{3,0}^{\text{(1)}}}{\widehat{\mathcal{W}}_{1,0}^{(1)}}=0.00119862\,,\\
&\frac{\widehat{\mathcal{W}}_{0,1}^{\text{(2)}}}{\widehat{\mathcal{W}}_{1,0}^{(2)}}=-0.161194,~\frac{\widehat{\mathcal{W}}_{0,2}^{\text{(2)}}}{\widehat{\mathcal{W}}_{1,0}^{(2)}}=0.00137037,~\frac{\widehat{\mathcal{W}}_{1,1}^{\text{(2)}}}{\widehat{\mathcal{W}}_{1,0}^{(2)}}=-0.00892794,~\frac{\widehat{\mathcal{W}}_{2,0}^{\text{(2)}}}{\widehat{\mathcal{W}}_{1,0}^{(2)}}=0.0158325,~\frac{\widehat{\mathcal{W}}_{3,0}^{\text{(2)}}}{\widehat{\mathcal{W}}_{1,0}^{(2)}}=0.000818889\,.
\end{split}
\ee
\end{scriptsize}
We can easily see that all of these coefficients satisfy the bounds listed in table \eqref{tab:boundsWkn2}, \eqref{tab:boundsWk}. We can compute the $\mathcal{C}_{p,q}$ using the equation \eqref{eq:CpqtoWk}. Some of them are
\be
\begin{split}
&\mathcal{C}_{0,4}=-0.0679822,~\mathcal{C}_{0,5}=-0.000596033,~\mathcal{C}_{0,6}=-0.000733546,~\mathcal{C}_{0,7}=-0.0000507897,\\
&\mathcal{C}_{1,1}=0.0551134,~\mathcal{C}_{1,2}=0.0864709,~\mathcal{C}_{1,3}=-0.0095459,~\mathcal{C}_{1,4}=0.00077458,~\mathcal{C}_{2,0}=0.15424,~\\
&\mathcal{C}_{2,1}=0.0264102,~\mathcal{C}_{2,2}=0.00454459,~\mathcal{C}_{3,0}=0.00251334\,.
\end{split}
\ee
Note that in equation \eqref{eq:CpqtoWk} we have used values in equations  \eqref{eq:WChiPT} with our normalizations $\wmW^{(k)}_{1,0}=1$. All of these $\mathcal{C}_{p,q}$ satisfy the two sided bounds presented in table \eqref{tab:Cpqbounds}.

\subsection{The $O(3)$ Lovelace-Shapiro model}
We demonstrate our bounds are satisfied by the $O(3)$ Lovelace-Shapiro model  eq \eqref{LS_model_full} with $\b_0=\b'=1/2$. Using the isospin amplitudes of the Lovelace-Shapiro model \eqref{LS_isoM}, one can compute the $\wmW^{(k)}_{p,q}$. Our bounds assumed the lower limit of the $s_1'$ integral as $\frac{8}{3}$ (for $\m=4$), while for the $O(3)$ Lovelace-Shapiro model lower limit is $s_1'=1$. Therefore, in order to compare with the bounds given in \eqref{tab:boundsWk}, we need to multiply with appropriate powers of $\frac{3}{8}$. In order to match with the conventions of the EFT scale, we multiply $\frac{\wmW^{(k)}_{n-m,m}}{\wmW^{(k)}_{1,0}}$ by $\left(\frac{3}{8}\right)^{m+2 n-2}$ (see \cite{PRAS,chout})
\begin{footnotesize}
\be
\begin{split}
&\frac{3 \widehat{\mathcal{W}}_{0,1}^{\text{(0)}}}{8 \widehat{\mathcal{W}}_{1,0}^{(0)}}=-0.472497,~\frac{9 \widehat{\mathcal{W}}_{2,0}^{\text{(0)}}}{64 \widehat{\mathcal{W}}_{1,0}^{(0)}}=0.130058,~\frac{27 \widehat{\mathcal{W}}_{1,1}^{\text{(0)}}}{512 \widehat{\mathcal{W}}_{1,0}^{(0)}}=-0.120704\,,\\
&\frac{3 \widehat{\mathcal{W}}_{0,1}^{\text{(1)}}}{8 \widehat{\mathcal{W}}_{1,0}^{(1)}}=-0.661605,~\frac{9 \widehat{\mathcal{W}}_{2,0}^{\text{(1)}}}{64 \widehat{\mathcal{W}}_{1,0}^{(1)}}=0.135717,~\frac{27 \widehat{\mathcal{W}}_{1,1}^{\text{(1)}}}{512 \widehat{\mathcal{W}}_{1,0}^{(1)}}=-0.145721\,,\\
&\frac{3 \widehat{\mathcal{W}}_{0,1}^{\text{(2)}}}{8 \widehat{\mathcal{W}}_{1,0}^{(2)}}=-0.385458,~\frac{9 \widehat{\mathcal{W}}_{2,0}^{\text{(2)}}}{64 \widehat{\mathcal{W}}_{1,0}^{(2)}}=0.13871,~\frac{27 \widehat{\mathcal{W}}_{1,1}^{\text{(2)}}}{512 \widehat{\mathcal{W}}_{1,0}^{(2)}}=-0.106502\,.
\end{split}
\ee
\end{footnotesize}
We can easily see that the RHS of the above equation all satisfy the two sided bounds presented in table \eqref{tab:boundsWkn2}, \eqref{tab:boundsWk}. Using the above equation and replacing $\frac{\wmW^{(k)}_{n-m,m}}{\wmW^{(k)}_{1,0}}\to \left(\frac{3}{8}\right)^{m+2 n-2}\frac{\wmW^{(k)}_{n-m,m}}{\wmW^{(k)}_{1,0}}$, we can calculate $\mathcal{C}_{p,q}$ from \eqref{eq:CpqtoWk}. With our normalization $\wmW^{(k)}_{1,0}=1$, we find
\be
\begin{split}
&\mathcal{C}_{0,4}=-0.0307213,~\mathcal{C}_{0,5}=-0.00188633,~\mathcal{C}_{1,1}=0.157499,~\mathcal{C}_{1,2}=0.0977439,\\
&\mathcal{C}_{1,3}=0.0111333,~\mathcal{C}_{2,0}=0.191501,~\mathcal{C}_{2,1}=0.0890061\,.
\end{split}
\ee
Notice that these $\mathcal{C}_{p,q}$ satisfy the inequalities listed in table \eqref{tab:Cpqbounds}.

\end{document}